\documentclass[aps,prd,superscriptaddress,10pt,nofootinbibt]{revtex4-2}

\usepackage{graphicx}
\usepackage{amsmath,amssymb}
\usepackage{hyperref}
\usepackage{xcolor}
\usepackage{caption}
\usepackage{subcaption}
\begin{document}

	\title{Anisotropic power-law inflation for models of non-canonical scalar fields non-minimally coupled to a two-form field}
	
	\author{Tuyen M. Pham}
	\email{tuyen.phammanh@phenikaa-uni.edu.vn}
	\affiliation{Phenikaa Institute for Advanced Study, Phenikaa University, Hanoi 12116, Vietnam}
	
	\author{Duy H. Nguyen}
	\email{duy.nguyenhoang@phenikaa-uni.edu.vn}
	\affiliation{Phenikaa Institute for Advanced Study, Phenikaa University, Hanoi 12116, Vietnam}
	
	\author{Tuan Q. Do }
	\email{tuan.doquoc@phenikaa-uni.edu.vn}
	\affiliation{Phenikaa Institute for Advanced Study, Phenikaa University, Hanoi 12116, Vietnam}
	\affiliation{Faculty of Basic Sciences, Phenikaa University, Hanoi 12116, Vietnam}

	\author{W. F. Kao}
	\email{homegore09@nycu.edu.tw}
	\affiliation{Institute of Physics, National Yang Ming Chiao Tung University, Hsin Chu 30010, Taiwan}
	\date{\today}
	
	\begin{abstract}
		In this paper, we investigate the validity of the so-called cosmic no-hair conjecture in the framework of anisotropic inflation models of non-canonical scalar fields non-minimally coupled to a two-form field. In particular, we focus on two typical {\it k}-inflation and Dirac-Born-Infeld inflation models, in which we find a set of exact anisotropic power-law inflationary solutions. Interestingly, these solutions are shown to be stable and attractive during an inflationary phase using the dynamical system analysis. The obtained results indicate that the non-minimal coupling between the scalar and two-form fields acts as a non-trivial source of generating stable spatial anisotropies during the inflationary phase and therefore violates the prediction of the cosmic no-hair conjecture, even when the scalar field is of non-canonical forms. In connection with the Planck 2018 data, tensor-to-scalar ratios of these anisotropic solutions are investigated.  As a result, it appears that the tensor-to-scalar ratio of the anisotropic power-law inflationary solution of {\it k}-inflation model turns out to be more highly consistent with the Planck 2018 data than that of Dirac-Born-Infeld  model.
		
	\end{abstract}
	
	
	\maketitle
	
	\section{Introduction} \label{sec1}
	Cosmic inflation  was firstly introduced as a solution to the longstanding puzzles in the standard Big Bang cosmology, such as the  horizon, flatness, and primordial monopole problems \cite{Starobinsky:1980te, Guth:1980zm, Linde:1981mu}. More interestingly, a rapid expansion during the inflationary phase will stretch the primordial density perturbations, which are created from quantum fluctuations, from microscopic scales to galactic scales, and therefore the large-scale structure of the present universe can be produced from these primordial density perturbations \cite{Lyth:2009zz}. For a recent interesting review on cosmic inflation, see Ref. \cite{Odintsov:2023weg}.
	
	Notably, observational data of the cosmic microwave background radiation (CMB) from the Wilkinson Microwave Anisotropy Probe (WMAP) \cite{WMAP:2012nax} and the Planck satellite \cite{Planck:2018vyg,Planck:2018jri} have been consistent very well with theoretical predictions of the standard inflationary models, whose underlying assumption is the cosmological principle, which states that our universe is spatially homogeneous and isotropic on large scales. In cosmology, there exists a unique spacetime called the Friedmann-Lemaitre-Robertson-Walker (FLRW) metric, which has both spatial homogeneity and isotropy. This is a reason why the FLRW metric has been widely used as a background spacetime in standard inflationary models \cite{Buchert:2015wwr}.
	
	However, some unavoidable anomalies in the CMB map, such as the cold spot and hemispheric asymmetry, have presented challenges to the standard inflationary models based on the cosmological principle \cite{Schwarz:2015cma}. In other words, if the cosmological principle is valid during the inflationary phase, the probability of the existence of the CMB anomalies is very small. It is worth noting that some other interesting observational evidences against the validity of cosmological principle have been summarized in a recent interesting review \cite{Aluri:2022hzs}. Therefore, violating the cosmological principle during the inflationary phase might be a reasonable resolution.  And one of the simplest ways to archive this violation is replacing the standard FLRW spacetime with the Bianchi spacetimes, which are spatially homogeneous but anisotropic \cite{Ellis:1968vb}. Consequently, we will end up with an anisotropic inflation \cite{Schwarz:2015cma}. Very interestingly, many theoretical predictions for anisotropic inflation had been worked out  even when the observed anomalies were not detected \cite{Pitrou:2008gk,Gumrukcuoglu:2007bx}.  It turns out that anisotropic inflation has been considered seriously by many people, e.g., see Ref.  \cite{Nojiri:2022idp} and references therein.
	
	Besides the cosmological principle, there has existed the so-called cosmic no-hair conjecture proposed by Hawking et al. a long time ago, which is also about the homogeneity and isotropy of universe's spacetime \cite{Gibbons:1977mu}. Basically, this conjecture implies that the late-time state of any accelerating universe is independent of its initial state. Eventually, the universe becomes isotropic and homogeneous as an attractor of cosmic evolution. Physicists and cosmologists  have long been challenged by the cosmic no-hair conjecture. However, they have made notable progress in providing partial proofs for this conjecture. Remarkably, the first  proof, specifically for the Bianchi spacetimes, which are homogeneous but anisotropic metrics, in the presence of a positive cosmological constant $\Lambda$, has been proposed by Wald using an approach based on energy conditions \cite{Wald:1983ky}. It should be noted that the Wald's proof deals only with background anisotropies. Remarkably, Starobinsky pointed out independently that the cosmological constant is the best ``isotropizer", i.e., it is capable of eliminating or extending over very large scales of all types of inhomogeneities \cite{Starobinsky:1982mr}. Furthermore, he concluded that the cosmic no-hair conjecture should be valid locally, i.e., inside the future de Sitter event horizon. More interestingly, an inhomogeneous time-independent tensor (we can regard it as hairs) existing outside of the future de Sitter event horizon may have an arbitrarily large amplitude.  In order to archive this conclusion, Starobinsky considered an inhomogeneous metric, whose scale factors depend on not only a cosmic time but also three spatial coordinates. This metric is nothing but an extension of de Sitter one with small inhomogeneous perturbations.  In a follow-up study, Starobinsky and his colleagues showed that inhomogeneous time-independent hairs also exist in a power-law inflation model of scalar field \cite{Muller:1989rp}. It is worth noting that similar conclusions have been archived later by other people in different scenarios, in which background spacetimes are of exact anisotropic and/or inhomogeneous forms like the so-called Tolman-Bondi spacetime \cite{Barrow:1984zz,Jensen:1986nf,SteinSchabes:1986sy}. According to these papers, one could state that the cosmic no-hair conjecture is invalid globally in the view of inhomogeneous hairs (a.k.a. constant perturbation modes in the super-Hubble regime) and may therefore be valid locally. Interestingly, these hairs could even be observable after the end of inflation and we could expect to see them directly in fluctuations of CMB temperature and polarization. 
	
	Recently, various cosmological models have been employed to test the validity of the cosmic no-hair conjecture, such that the higher curvature models \cite{Barrow:2005qv,Middleton:2010bv}, the Lorentz Chern-Simons theory \cite{Kaloper:1991rw}, and the Horndeski models \cite{Tahara:2018orv}. It turns out that some claimed counterexamples from these models no longer hold due to their instability during the inflationary phase as indicated in Ref. \cite{Kao:2009zza}. From 2009 to 2010, a vivid counterexample to the cosmic no-hair conjecture was successfully constructed by Kanno, Soda, and Watanabe (KSW) within a supergravity-motivated model involving a non-minimal coupling between the $U(1)$ gauge and scalar fields such as $f^2(\phi)F^{\mu\nu} F_{\mu\nu}$ \cite{Watanabe:2009ct,Kanno:2010nr}. As a result, the KSW model admits homogeneous and anisotropic Bianchi type I solutions, whose spatial anisotropies turn out to be stable during the inflationary phase, in contrast to the prediction of the cosmic no-hair conjecture. Subsequently, several non-trivial extensions of the KSW model have been proposed by considering non-canonical scalar fields instead of canonical one, such as the Dirac-Born-Infeld (DBI) model \cite{Do:2011zz,Ohashi:2013pca,Holland:2017cza,Nguyen:2021emx}, the generalized ghost condensate model \cite{Ohashi:2013pca}, supersymmetric DBI model \cite{Do:2016ofi}, and the {\it k}-inflation model \cite{Do:2020hjf}. As a result, the cosmic no-hair conjecture is always violated in these non-canonical extensions. For CMB imprints of non-canonical anisotropic inflation, see Refs. \cite{Do:2016ofi,Do:2020hjf,Do:2020ler}, while that of canonical one can be found in Refs. \cite{Watanabe:2010fh,Dulaney:2010sq,Gumrukcuoglu:2010yc,Watanabe:2010bu,Bartolo:2012sd,Chen:2014eua}. All these results indicate that the non-minimal coupling, $f^2(\phi)F^{\mu\nu}F_{\mu\nu}$, has played a leading role in order to generate stable spatial anisotropies during the inflationary phase.  Interesting reviews on cosmological implications of the anisotropic inflation based on the vector field can be found in Refs. \cite{Soda:2012zm,Maleknejad:2012fw}.
	
	One might ask if spatial anisotropies can also be caused by other mechanisms. Interestingly,  people have figured out that a non-minimal coupling between scalar and two-form  fields such as $f^2(\phi)H^{\mu\nu\rho}H_{\mu\nu\rho}$ could play a similar role as the coupling $f^2(\phi)F^{\mu\nu}F_{\mu\nu}$    \cite{Ohashi:2013mka,Ohashi:2013qba,Ito:2015sxj,Do:2018zac,Almeida:2019xzt}.   It is well known that a two-form field $B_{\mu\nu}$ can be found in string theory \cite{Lidsey:1999mc}. In Ref. \cite{Ohashi:2013mka}, the authors have pointed out that  the anisotropy induced by the two-form field corresponds to the prolate type, i.e., the expansion of the Universe slows down in the $(y, z)$ plane, in contrast to the
	oblate type stemming from the vector field (a.k.a. one-form field). In Ref. \cite{Ohashi:2013qba}, the authors have derived the corresponding observational constraints of two anisotropic inflation models, one for the vector field and the other for the two-form field. Very interestingly, some significant gaps between observational predictions of these two types of anisotropic inflation have been identified accordingly. The authors have come to the main conclusion that the precise measurements of $g_\ast$ as well as the TB correlation will clarify which anisotropic model is favored over the other \cite{Ohashi:2013qba}.  In Ref. \cite{Ito:2015sxj}, the authors have investigated anisotropic hairs in the presence of both one-form and two-form fields. As a result, they have obtained an important conclusion that there always exists one stable anisotropic fixed point in this model. In Ref. \cite{Do:2018zac}, the authors have indicated that the five-dimensional (5D) two-form field can be shown to be equivalent to a 5D gauge field via a Routh transformation. Interestingly, the cosmic no-hair conjecture has been shown to be broken down in this 5D model. In Ref. \cite{Almeida:2019xzt}, the authors have considered a more general scenario, which involves not only one- and two-form fields but also a three-form one. It should be noted that all these studies of anisotropic inflation in the presence of the two-form field have only been done for the canonical field.  Other physical and cosmological implications of two-form field can be found in Refs. \cite{Kao:1992xz,Kao:1996ea,Elizalde:2018now,Elizalde:2018rmz,BeltranAlmeida:2019fou,Do:2020ojg,Almeida:2020lsn,Paul:2020bdy,Paul:2020duu,Paul:2022mup,Fujita:2022ait}. 
	
	Motivated by studies of anisotropic inflation models of non-canonical scalar fields \cite{Do:2011zz,Ohashi:2013pca,Do:2016ofi,Do:2020hjf} and anisotropic inflation models of two-form fields \cite{Ohashi:2013qba,Ohashi:2013mka,Ito:2015sxj,Do:2018zac,Almeida:2019xzt}, we would like to investigate in the present paper the validity of the cosmic no-hair conjecture in a mixed scenario, in which a non-canonical scalar field is allowed to non-minimally coupled to the two-form field. Specifically, we focus on two well-known types of non-canonical scalar fields, one is from the {\it k}-inflation model \cite{Armendariz-Picon:1999hyi,Garriga:1999vw} and the other is from the string-inspired DBI inflation model \cite{Silverstein:2003hf,Baumann:2006cd,Copeland:2010jt}. As a result, we are able to find exact anisotropic power-law solutions for both non-canonical scalar fields. Furthermore, we confirm, using the dynamical system method, that these solutions are indeed stable and attractive during the inflationary phase and therefore act as additional counterexamples to the cosmic no-hair conjecture. In connection with the Planck 2018 data \cite{Planck:2018vyg,Planck:2018jri} as well as future detections like the CMB-S4 project \cite{CMB-S4:2020lpa}, we will derive a tensor-to-scalar ratio for a general case of non-canonical scalar field non-minimally coupled to the two-form field. Then we will focus on the two mentioned models, i.e., {\it k}-inflation and DBI ones, to see whether the corresponding ratios are consistent with the Planck 2018 data or not. Very interestingly, the obtained tensor-to-scalar ratio of the {\it k}-inflation two-form field model turns out to be highly consistent with the Planck 2018 data.
	
	As a result, this paper will be organized as follows: (i) An introduction of our study has been written in Sect. \ref{sec1}. (ii) A general action of studied models will be presented in Sect. \ref{sec2}. (iii) In Sect. \ref{sec3}, we derive the corresponding set of anisotropic power-law solutions for {\it k}-inflation model and investigate its stability. (iv) In Sect. \ref{sec4}, we extend our analysis to the DBI model.  (v) In Sect. \ref{sec5}, we compare the anisotropic parameter $|\Sigma/H|$ derived in this paper with that obtained in the previous papers for heuristic reasons.  (vi) In Sect. \ref{sec6}, tensor-to-scalar ratios of the obtained anisotropic power-law inflationary solutions will be investigated. (vii) Finally, concluding remarks will be written in Sect. \ref{final}. Additional calculations will be presented in the Appendix.
	\section{General action} \label{sec2}
	Let us begin by introducing a general action of non-canonical extension of the KSW model \cite{Ohashi:2013pca,Do:2020ler},
	\begin{equation}\label{2-1}
		S = \int d^4x\sqrt{-g}\left[\dfrac{R}{2}+P(\phi,X) - \dfrac{1}{4}f^2(\phi)F_{\mu\nu}F^{\mu\nu}\right],
	\end{equation}
	where $P(\phi,X)$ is an arbitrary function of scalar field $\phi$ and its kinetic term defined as $X \equiv -(1/2) \partial_\mu \phi \partial^\mu \phi$ \cite{Armendariz-Picon:1999hyi}. In addition, $f(\phi)$ is the gauge kinetic function depending only on $\phi$, while the rank-2 tensor $F_{\mu\nu} = \partial_\mu A_\nu - \partial_\nu A_\mu$ is the field strength of the vector field $A_\mu$. It is noted that we have set the reduced Planck mass as one, i.e., $M_{\text{p}} = 1$, just for convenience. Specific forms of $P(\phi,X)$ have been considered in previous papers \cite{Do:2011zz,Ohashi:2013pca,Do:2016ofi,Do:2020hjf}.
	
	In this paper, we would like to investigate a modification of the above action, in which the one-form field $A_\mu$ is replaced by a two-form field $B_{\mu\nu}$. As a result, the corresponding action is given by 
	\begin{equation}\label{2-3}
		S = \int d^4x\sqrt{-g}\left[\dfrac{R}{2}+P(\phi,X) - \dfrac{1}{12}f^2(\phi)H_{\mu\nu\rho}H^{\mu\nu\rho}\right],
	\end{equation}
	where
	\begin{equation}\label{2-2}
		H_{\mu\nu\rho} = \partial_\mu B_{\nu\rho}+\partial_\nu B_{\rho\mu}+\partial_\rho B_{\mu\nu}.
	\end{equation}
	Without the loss of generality, one can assume that the $(y, z)$ plane aligns with the direction of the two-form field. Consequently, we can express $B_{\mu\nu}$ in the following form  \cite{Ohashi:2013qba,Ohashi:2013mka,Ito:2015sxj}
	\begin{equation}\label{2-4}
		\frac{1}{2}B_{\mu\nu}dx^\mu \wedge dx^\nu = v_B(t) dy\wedge dz,
	\end{equation}
	where $v_B(t)$ is a function of cosmic time $t$. In this paper, we will study action \eqref{2-3} for two typical non-canonical scalar fields, one is from the {\it k}-inflation model \cite{Do:2020hjf,Armendariz-Picon:1999hyi} and the other is from the string-inspired DBI inflation model \cite{Do:2011zz,Silverstein:2003hf,Baumann:2006cd,Copeland:2010jt}.
	\section{ $K$-inflation case} \label{sec3}
	In this section, we focus on the {\it k}-inflation model \cite{Do:2020hjf,Armendariz-Picon:1999hyi}, in which the function $P(\phi,X)$ takes the following form $P(\phi, X) = K(\phi)X + L(\phi)X^2$. Thus, the action \eqref{2-3} becomes
	\begin{equation}\label{3-1}
		S = \int d^4x\sqrt{-g}\left[\dfrac{R}{2}+K(\phi)X + L(\phi)X^2 - \dfrac{1}{12}f^2(\phi)H_{\mu\nu\rho}H^{\mu\nu\rho}\right],
	\end{equation}
	here $K(\phi)$ and $L(\phi)$ are arbitrary functions of $\phi$.
	Following Refs. \cite{Ohashi:2013mka,Ohashi:2013qba,Ito:2015sxj}, we will adopt the Bianchi type I metric,
	\begin{equation}\label{3-2}
		ds^2  = -N^2(t) dt^2+e^{2\alpha(t)}[e^{-4\sigma(t)}dx^2+e^{2\sigma(t)}(dy^2+dz^2)],
	\end{equation}
	as the background spacetime for the cosmic evolution.
	Here, $N(t)$ represents the lapse function that allows us to derive the Friedmann constraint equation, while $\alpha$ represents the average expansion measured in terms of the number of e-foldings and $\sigma$ corresponds to the spatial anisotropy.
	By substituting the background metric into the action \eqref{3-1}, we obtain the following expression,
	\begin{equation}\label{3-3}
		S = \int d^4 x e^{3\alpha}\left[3\dfrac{(\dot{\sigma}^2 - \dot{\alpha}^2)}{N}+\dfrac{K(\phi)}{2 N}\dot{\phi}^2+\dfrac{L(\phi)}{4 N^3}\dot{\phi}^4+\dfrac{f^2(\phi)}{2 N}e^{-4(\alpha+\sigma)}\dot{v}_B^2\right],
	\end{equation}
	where an overdot stands for a derivative with respect to the cosmic time $t$.
	The equation of motion for the two-form field can be solved to give a non-trivial solution of $v_B(t)$ as 
	\begin{equation}\label{3-4}
		\dot{v}_B = p_B f^{-2}e^{\alpha+4\sigma},
	\end{equation}
	where $p_B$ is a constant of integration \cite{Ohashi:2013mka,Ohashi:2013qba,Ito:2015sxj}. By varying the above action \eqref{3-3} with respect to $ N$, $\alpha$, $\sigma$, as well as $\phi$,  we are able to derive the following background equations of motion,
	\begin{align}\label{3-5}
		\dot{\alpha }^2&=\dot{\sigma }^2+\frac{\dot{\phi }^2}{6}  K+\frac{\dot{\phi }^4}{4}  L+\frac{e^{4 \sigma -2 \alpha } p_B^2}{6 f^2},\\\label{3-6}
		\ddot{\alpha }&=-3 \dot{\alpha }^2+\frac{\dot{\phi }^4}{4}  L+\frac{e^{4 \sigma -2 \alpha } p_B^2}{3 f^2},\\\label{3-7}
		\ddot{\sigma }&=-3 \dot{\alpha } \dot{\sigma }-\frac{e^{4 \sigma -2 \alpha } p_B^2}{3 f^2},\\\label{3-8}
		\ddot{\phi } \left(K + 3 \dot{\phi }^2 L\right)&=-3 \dot{\alpha } \dot{\phi } K- 3 \dot{\alpha } \dot{\phi }^3 L-\frac{1}{2} \dot{\phi }^2 K_{\phi }-\frac{3}{4} \dot{\phi }^4 L_{\phi }+\frac{e^{4 \sigma -2 \alpha } p_B^2 f_{\phi }}{f^3},
	\end{align}
	after setting $N=1$, respectively. Here, the subscript in $K_\phi, L_\phi $ and $f_\phi$ indicates a derivative with respect to the field $\phi$, i.e., $K_\phi \equiv \partial K/\partial \phi$. To figure out analytical solutions for these derived field equations, we will consider the following ansatz as used in many previous papers \cite{Kanno:2010nr,Do:2011zz,Ohashi:2013mka,Ohashi:2013qba,Ito:2015sxj,Do:2020hjf},
	\begin{equation}\label{3-9}
		\alpha = \zeta \log{(t)};\quad\sigma = \eta \log{(t)};\quad\phi = \xi \log{(t)}+\phi_0,
	\end{equation}
	along with the exponential functions of the scalar field,
	\begin{align}\label{3-10}
		K(\phi) &= k_0 e^{\kappa \phi},\\
		L(\phi) &= l_0 e^{\lambda\phi},\\
		f(\phi) &= f_0 e^{\rho\phi},
	\end{align}
	where $k_0$, $l_0$, $f_0$, $\zeta$, $\eta$, $\xi$, $\phi_0$, $\lambda$, $\kappa$, and $\rho$ are all constant. For convenience, we will introduce the following new parameters to aid in our analysis,
	\begin{align}\label{3-11}
		u &= l_0 \exp[\lambda \phi_0],\\
		w &= p_B^2 f_0^{-2} \exp[-2\rho \phi_0].
	\end{align}
	Thus, the field Eqs. \eqref{3-5}, \eqref{3-6}, \eqref{3-7}, and \eqref{3-8} can be simplified to the following algebraic equations,
	\begin{align}\label{19}
		-\zeta ^2+\eta ^2+\frac{k_0 \xi ^2}{6}+\frac{\xi ^4 u}{4}+\frac{w}{6}&=0,\\\label{20}
		3 \zeta ^2-\zeta -\frac{\xi ^4 u}{4}-\frac{w}{3}&=0,\\\label{21}
		3 \zeta  \eta -\eta +\frac{w}{3}&=0,\\\label{22}
		3 \zeta  k_0 \xi -k_0 \xi +3 \zeta  \xi ^3 u-\frac{3 \xi ^3 u}{2}-\rho w&=0.
	\end{align}
	In addition to these equations, we have the following constraint equations that ensure all terms in the field equations are proportional to $t^{-2}$,
	\begin{align}\label{23}
		\lambda \xi &=2,\\\label{24}
		\kappa&=0,\\\label{25}
		-2 \zeta +4 \eta -2 \xi  \rho &= -2.
	\end{align}
	Apparently, the condition for $\kappa =0$ gives us a constant value of $K(\phi) = k_0$ for the ansatz \eqref{3-9}.
	
	\subsection{Anisotropic power-law inflation} \label{sec2.2}
	We will focus on seeking anisotropic solutions. From Eq. \eqref{25} we have
	\begin{equation}\label{33}
		\eta =\frac{\zeta }{2}+\frac{\xi  \rho }{2}-\frac{1}{2}.
	\end{equation}
	On the other hand, we have from  Eqs. \eqref{21} and \eqref{22} that
	\begin{align}
		u&= -\frac{\lambda ^2 \left(9 \zeta ^2 \lambda  \rho -12 \zeta  \lambda  \rho +18 \zeta  \rho ^2+12 \zeta  k_0+3 \lambda  \rho -4 k_0-6 \rho ^2\right)}{24 (2 \zeta -1)},\\
		w&=-\frac{3 \left(3 \zeta ^2 \lambda -4 \zeta  \lambda +6 \zeta  \rho +\lambda -2 \rho \right)}{2 \lambda },
	\end{align}
	thanks to the constraint equation \eqref{23} as well as the relation shown in Eq. \eqref{33}.
	Substituting $\eta$, $u$, and $w$ defined above into either Eq. \eqref{19} or Eq. \eqref{20}, we can find non-trivial solutions of $\zeta$ such as
	\begin{align}
		\zeta_{\pm} = \frac{5}{12}-\frac{5 \rho }{12 \lambda }\pm\frac{\sqrt{(\lambda -\rho ) (\lambda +23 \rho )-32 k_0}}{12 \lambda }.
	\end{align}
	Now, we rewrite Eq. \eqref{33} as follows
	\begin{equation} \label{key-0}
		\zeta = 2 \eta -\frac{2 \rho }{\lambda }+1.
	\end{equation}
	This equation implies an important point that the ratio $-2\rho/\lambda$ will mainly determine the value of $\zeta$ since $\eta$ should be much smaller than $\zeta$. Therefore, the constraint for the existence of inflation, $\zeta\gg 1$, will imply that $\zeta\simeq -{2\rho}/{\lambda}\gg1$, or equivalently $|\rho| \gg |\lambda|$. Consequently, we observe that only the solution,
	\begin{equation}\label{key-1}
		\zeta= \zeta_{-} = \frac{5}{12}-\frac{5 \rho }{12 \lambda }-\frac{\sqrt{(\lambda -\rho ) (\lambda +23 \rho )-32 k_0}}{12 \lambda },
	\end{equation}
	is suitable for describing the inflationary phase.
	Note that the positivity of $\rho$ will imply the negativity of $\lambda$, or vice versa. In this paper, we will prefer the choice, in which $\rho$ is assumed to be positive and $\lambda$ will be negative.  Then, the corresponding expression for $\eta$ turns out to be
	\begin{equation}\label{key-2}
		\eta=-\frac{7}{24}+\frac{19 \rho }{24 \lambda }-\frac{\sqrt{(\lambda -\rho ) (\lambda +23 \rho )-32 k_0}}{24 \lambda },
	\end{equation}
	while the corresponding anisotropy parameter reads 
	\begin{equation}\label{key-3}
		\dfrac{\Sigma }{H}\equiv \dfrac{\dot{\sigma}}{\dot{\alpha}} = \dfrac{\eta}{\zeta_-} =\frac{8 k_0-3 (\lambda -2 \rho ) \left[3 \lambda -3 \rho +\sqrt{(\lambda -\rho ) (\lambda +23 \rho )-32 k_0} \right]}{12(\lambda -\rho ) (\lambda -2 \rho )+16k_0}.
	\end{equation}
	The real values of $\zeta$ and $\eta$ require that $k_0$ must satisfy the following constraint,
	\begin{equation}\label{key-4}
		k_0\leq\dfrac{(\lambda -\rho ) (\lambda +23 \rho )}{32} \simeq -\frac{23}{32}\rho^2.
	\end{equation}
	However, the smallness of $\eta$ implies, according to Eq. \eqref{key-2}, the following constraint,
	\begin{equation}\label{key-5}
		\frac{\sqrt{(\lambda -\rho ) (\lambda +23 \rho ) -32 k_0}}{24 \lambda }\simeq\dfrac{19\rho}{24\lambda},
	\end{equation}
	which leads to an approximated value of $k_0$ such as
	\begin{equation}\label{key-6}
		k_0\simeq -12 \rho^2,
	\end{equation}
	due to the constraint, $\zeta \simeq -2{\rho}/{\lambda}$. It is clear that this value is consistent with the inequality \eqref{key-4}. Consequently, the value for $\zeta$ and $\eta$ can be approximate as follows
	\begin{align}
		\zeta &\simeq \dfrac{5}{12}-2\dfrac{\rho}{\lambda}\simeq -2\dfrac{\rho}{\lambda},\\
		\eta&\simeq -\dfrac{6}{19} <0.
	\end{align}
	It is clear that the negativity of $\eta$ is consistent with the positivity of $w$, according to Eq. \eqref{21}. However,  the negativity of $\eta$ in this model of two-form field is indeed in contrast to the positivity of $\eta$ required in models of vector field \cite{Kanno:2010nr,Do:2011zz,Do:2020hjf}. This result is consistent with the previous investigation for canonical scalar field \cite{Ito:2015sxj}.
	
	As a result, the following approximated value of the anisotropy parameter turns out to be 
	\begin{equation}\label{key-7}
		\dfrac{\Sigma}{H}=\dfrac{\eta}{\zeta} \simeq \frac{3 \lambda }{19 \rho } <0.
	\end{equation}
	Of course the absolute value of this ratio is much smaller than one, i.e., $|\Sigma/H|\ll 1$ as expected.  In fact, to be consistent with the cosmological observation the absolute value of the anisotropy parameter, i.e., $|\Sigma/H|$, must be much smaller than one \cite{Watanabe:2009ct,Kanno:2010nr}.  For heuristic reasons, we will compare $|\Sigma/H| $ derived in our current model with that obtained in the KSW model \cite{Kanno:2010nr} and the {\it k}-inflation model \cite{Do:2020hjf} of vector field, as well as with that derived in Ref. \cite{Ito:2015sxj} for the canonical scalar field coupled to the two-form field, using specific values of field parameters such as $|\lambda|=0.1$ and $\rho=50$. As a result, we obtain the corresponding values as
	$|\Sigma/H  |_{\text {KSW}}\simeq 0.0004$, $|\Sigma/H |_{k\text {-one-form}}\simeq 0.0005$, $|\Sigma/H |_{\text{canonical-two-form}}\simeq 0.0008$, and $|\Sigma/H |_{k\text{-two-form}}\simeq 0.0003 <|\Sigma/H |_{\text{canonical-two-form}}$. It turns out that the anisotropy induced by the two-form field non-minimally coupled to the {\it k}-inflation field is the smallest one among these four values.
	
	\subsection{Stability analysis}
	To investigate the stability of the anisotropic power-law solution defined above
	we will introduce the corresponding dynamical variables  \cite{Ito:2015sxj},
	\begin{equation}\label{key-8}
		x \equiv \dfrac{\dot{\sigma}}{\dot{\alpha}},\quad y\equiv\dfrac{\dot{\phi}}{\dot{\alpha}},\quad z \equiv p_B \dfrac{ e^{-\alpha+2\sigma}}{f(\phi)\dot{\alpha}},
	\end{equation} 
	along with two auxiliary variables \cite{Do:2020hjf}, 
	\begin{align}
		\omega_\kappa \equiv e^{\kappa\phi/2},\quad\omega_\lambda\equiv\sqrt{l_0}\dot{\alpha}e^{\lambda\phi/2}.
	\end{align}
	As a result, the corresponding dynamical system of autonomous equations are defined to be
	\begin{align}
		\dfrac{dx}{d\alpha} &=\dfrac{\ddot{\sigma}}{\dot{\alpha}^2}-x\dfrac{\ddot{\alpha}}{\dot{\alpha}^2},\\
		\dfrac{dy}{d\alpha} &=\dfrac{\ddot{\phi}}{\dot{\alpha}^2}-y\dfrac{\ddot{\alpha}}{\dot{\alpha}^2},\\
		\dfrac{dz}{d\alpha}&= z (2 x-\rho  y-\dfrac{\ddot{\alpha}}{\dot{\alpha}^2} -1),\\
		\dfrac{d\omega_\lambda}{d\alpha}&=\dfrac{\lambda }{2} y \omega_\lambda+\dfrac{\ddot{\alpha}}{\dot{\alpha}^2}\omega_\lambda,\\
		\dfrac{d\omega_\kappa}{d\alpha} &=\dfrac{\kappa}{2} y \omega_\kappa.
	\end{align}
	Here, $\alpha =\int \dot\alpha dt$ is understood as a new time coordinate \cite{Kanno:2010nr,Do:2011zz,Ito:2015sxj,Do:2020hjf}. In the above equations, there exist $\ddot\alpha$, $\ddot\sigma$, and $\ddot\phi$, which can be determined from the field equations \eqref{3-6}, \eqref{3-7}, and \eqref{3-8} with the help of the Friedmann constraint equation \eqref{3-5}, which now becomes as
		\begin{equation}\label{key-9}
		\frac{1}{6} k_0 y^2 \omega _{\kappa }^2+x^2+\frac{1}{4} y^4 \omega _{\lambda }^2+\frac{z^2}{6}-1 = 0.
	\end{equation}
As a result, explicit expressions of autonomous equations can be defined to be
	\begin{align}\label{54}
		\dfrac{dx}{d\alpha}&=\frac{1}{6} k_0 \omega_\kappa^2 x y^2+x^3-\frac{x z^2}{6}-x-\frac{z^2}{3},\\\label{55}
		\dfrac{dy}{d\alpha}&=\frac{1}{4 \left(k_0 \omega _{\kappa }^2+3 y^2 \omega _{\lambda }^2\right)}\left[-\frac{2}{3} k_0 y \omega _{\kappa }^2 \left(-k_0 y^2 \omega _{\kappa }^2-6 x^2+z^2-12\right)\right.\nonumber\\
		&\qquad\qquad\qquad\qquad\quad\left.-2 y^3 \omega _{\lambda }^2 \left(-k_0 y^2 \omega _{\kappa }^2-6 x^2+z^2-12\right)-2 \kappa  k_0 y^2 \omega _{\kappa }^2\right.\nonumber\\
		&\qquad\qquad\qquad\qquad\quad\left.-12 k_0 y \omega _{\kappa }^2-3 \lambda  y^4 \omega _{\lambda }^2-12 y^3 \omega _{\lambda }^2+4 \rho  z^2\right],\\\label{56}
		\dfrac{dz}{d\alpha}&= z \left(\frac{1}{6} k_0 \omega_{\kappa }^2 y^2+x^2+2 x-\rho  y-\frac{z^2}{6}+1\right),\\\label{57}
		\frac{d\omega _{\lambda }}{\text{d$\alpha $}}&=-\omega _{\lambda } \left(\frac{k_0\omega_{\kappa }^2 y^2}{6}+x^2-\frac{z^2}{6}+2\right)+\frac{1}{2} \lambda  y \omega _{\lambda },\\\label{58}
		\dfrac{d\omega_\kappa}{d\alpha} &= \frac{1}{2} \kappa  y \omega _{\kappa }.	
	\end{align}
		Now, we are going to find out anisotropic fixed points with $x\neq 0$ to this dynamical system by solving the following equations,
	\begin{equation}\label{key-10}
		\dfrac{dx}{d\alpha} = \dfrac{dy}{d\alpha}=\dfrac{dz}{d\alpha} = \dfrac{d\omega_\lambda}{d\alpha}=\dfrac{d\omega_\kappa}{d\alpha} = 0.
	\end{equation}
	First, the equation $d\omega_\kappa/d\alpha = 0$ gives $\kappa = 0$ and then $\omega_\kappa=1$, consistent with the power-law solution derived in the previous subsection. As a result, the equation ${d\omega_\lambda}/{d\alpha} = 0$ gives
	\begin{equation}\label{key-11}
		\frac{k_0 y^2}{6}+x^2-\frac{\lambda  y}{2}-\frac{z^2}{6}+2=0,
	\end{equation}
	while the equation ${dz}/{d\alpha}= 0$
	implies
	\begin{equation}\label{key-12}
		\frac{k_0 y^2}{6}+x^2+2 x-\rho  y-\frac{z^2}{6}+1 = 0.
	\end{equation}
	Then, we can obtain the expression for $y$ from these two equations,
	\begin{equation}\label{key-13}
		y = \dfrac{2-4x}{\lambda-2 \rho}.
	\end{equation}
	On the other hand, by using two equations, ${dx}/{d\alpha} = 0$ and ${dz}/{d\alpha} = 0$, we can obtain the expression for $z^2$ as
	\begin{equation}\label{key-14}
		z^2 = -3 x (2 x-\rho  y+2).
	\end{equation}
	Thanks to these results, we are able to obtain non-trivial solutions for $x$ from either equation $dy/d\alpha=0$ or $dz/d\alpha=0$,
	\begin{equation}\label{key-15}
		x_{\pm} = \frac{-3 (\lambda -2 \rho ) \left[3 \lambda -3 \rho\pm\sqrt{(\lambda -\rho ) (\lambda +23 \rho )-32 k_0} \right]+8 k_0}{4 \left(3 (\lambda -\rho ) (\lambda -2 \rho )+4 k_0\right)}.
	\end{equation}
	It is clear that the solution $x_+$ is absolutely equivalent to the anisotropic power-law one found in the previous subsection due to the result $x_+=\eta/\zeta_-$. This indicates that the stability of the anisotropic fixed point is also that of the anisotropic power-law solution. 
	
\begin{figure}[htp!]
		\centering
		\includegraphics[scale=0.6]{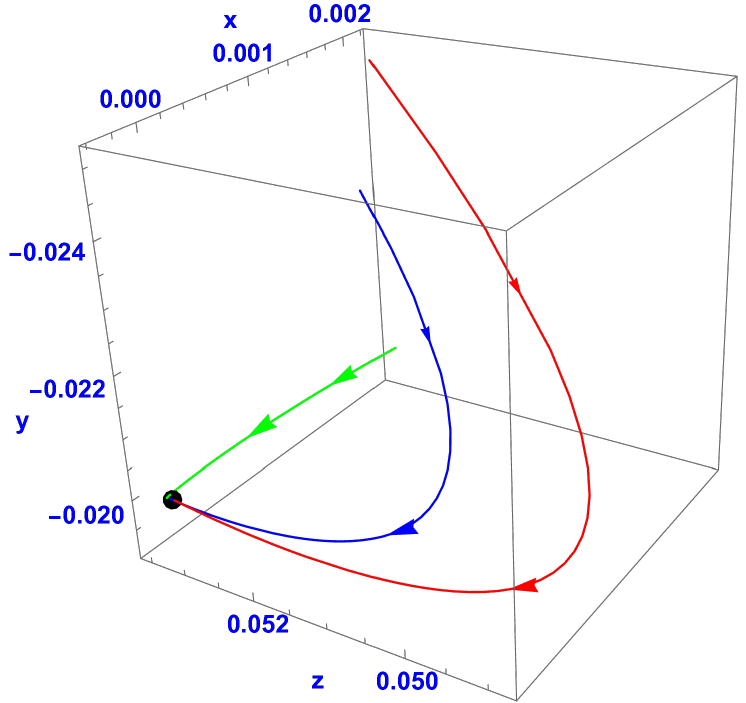}
		\caption{Numerical results demonstrate the attractive behavior of the obtained anisotropic fixed point (displayed as a black point). Here, the parameters have been chosen as $\lambda=-0.1$ and $\rho=50$. }
		\label{fig:numerical-k}
	\end{figure}

	As a result, approximated values for the anisotropic fixed point can be defined as
	\begin{align}
		x&\simeq \frac{3 \lambda }{19 \rho },\\
		y&\simeq -\dfrac{1}{\rho},\\
		z^2&\simeq -\frac{27\lambda }{19 \rho },\\
		\omega_\lambda^2&\simeq12 \rho ^4-\frac{94 \lambda  \rho ^3}{19},
	\end{align}
	using the constraints pointed out above, i.e., $\rho\gg |\lambda| $, $k_0\simeq -12\rho^2$, and $\zeta\simeq -2{\rho}/{\lambda}\gg 1$, for the inflationary phase.  To examine the stability of the anisotropic fixed point, we perturb the autonomous equations around this fixed point as follows
	\begin{align}
		\dfrac{d\delta x}{d\alpha} &\simeq -3\delta x+\frac{12\lambda }{19}\delta y-2 \sqrt{-\frac{3 \lambda }{19 \rho }}\delta z,\\
		\dfrac{d\delta y}{d\alpha} &\simeq -\frac{6 \lambda }{19 \rho ^2}\delta x-7\delta y+\frac{5}{4} \sqrt{\frac{-3 \lambda }{19 \rho ^3}} \delta z+\frac{\sqrt{3 }}{2\rho^3}\delta \omega_\lambda,\\
		\dfrac{d\delta z}{d\alpha} &\simeq 6 \sqrt{\frac{-3 \lambda }{19 \rho }}\delta x+9 \sqrt{\frac{-3\lambda\rho}{19}}\delta y+\ \frac{39 \lambda }{38 \rho }\delta z,\\
		\dfrac{d\delta\omega_\lambda}{d\alpha}&\simeq-\frac{12}{19} \sqrt{3} \lambda  \rho \delta x-8 \sqrt{3} \rho ^3\delta y+6 \sqrt{\frac{-\lambda}{19}  \rho ^3} \delta z-\frac{14 \lambda }{19 \rho }\delta \omega _{\lambda }.
	\end{align}
	Then, we take exponential perturbations,
	\begin{align}
		\delta x &= A_x \exp[\omega\alpha],\\
		\delta y &= A_y \exp[\omega\alpha],\\
		\delta z &= A_z \exp[\omega\alpha],\\
		\delta \omega_{\lambda } &= A_{\omega} \exp[\omega\alpha],
	\end{align}
	to obtain the corresponding matrix equation,
	\begin{align}
		\mathcal{M}  \begin{pmatrix}
			A_x\\
			A_y\\
			A_{z}\\
			A_{\omega}
		\end{pmatrix}&\equiv	
		\left(
		\begin{array}{cccc}
			-\omega -3 & \frac{12 \lambda }{19} & -2 \sqrt{\frac{-3 \lambda }{19 \rho }} & 0 \\
			-\frac{6 \lambda }{19 \rho ^2} & -\omega -7 & \frac{5}{4} \sqrt{\frac{-3 \lambda }{19 \rho ^3}} & \frac{\sqrt{3}}{2 \rho ^3} \\
			6 \sqrt{\frac{-3 \lambda }{19 \rho }} & 9 \sqrt{\frac{-3\lambda\rho}{19} } & -\omega+\frac{39 \lambda }{38 \rho }  & 0 \\
			-\frac{12}{19} \sqrt{3} \lambda  \rho  & -8 \sqrt{3} \rho ^3 & 6 \sqrt{\frac{-\lambda}{19}  \rho ^3} & -\omega-\frac{14 \lambda }{19 \rho } \\
		\end{array}
		\right)
		\begin{pmatrix}
			A_x\\
			A_y\\
			A_{z}\\
			A_{\omega}
		\end{pmatrix}=0.
	\end{align}
	It is well known that this matrix equation admits non-trivial solutions if and only if
	\begin{equation}\label{key-16}
		\det \mathcal{M} = 0,
	\end{equation}
	which can be expanded to be a polynomial equation for $\omega$ as follows
	\begin{equation}\label{80}
		a_4 \omega^4+a_3\omega^3+a_2\omega^2+a_1\omega+a_0=0,
	\end{equation}
	where
	\begin{align}
		a_4 & =1 >0,\\
		a_3 &\simeq 10-\frac{11 \lambda }{38 \rho },\\
		a_2 &\simeq  33-\frac{229 \lambda }{76 \rho },\\
		a_1&\simeq36-\frac{1677 \lambda }{76 \rho },\\
		a_0&\simeq -\frac{891 \lambda }{19 \rho }.
	\end{align}
	Apparently, the constraint that $\lambda<0$ and $\rho>0$ will indicate the fact that all the coefficients $a_i$ $(i=0-3)$ are positive. Hence, the polynomial equation \eqref{80} only admits negative roots $\omega<0$. Therefore, the anisotropic fixed point turns out to be stable against perturbations. In addition, the numerical result shown in Fig. \ref{fig:numerical-k}  confirms that the anisotropic fixed point is clearly attractive since  trajectories with different initial conditions all converge to the anisotropic fixed point (displayed as the black point) rather than an isotropic fixed point with $x=z=0$. For now, we can conclude that the anisotropic power-law solution is stable and attractive during the inflationary phase. And this model provides us one more  counterexample to the cosmic no-hair theorem.
	\section{Dirac-Born-Infeld case} \label{sec4}
	In this section, we extend our analysis to the DBI model, whose origin comes from string theory \cite{Silverstein:2003hf,Baumann:2006cd,Copeland:2010jt}. Anisotropic power-law inflation for the DBI model in the presence of one-form field has been firstly studied in Ref. \cite{Do:2011zz} then revisited in Refs. \cite{Ohashi:2013pca,Holland:2017cza,Nguyen:2021emx}. An action of the Dirac-Born-Infeld model in the presence of the two-form field is given by
	\begin{equation}\label{86}
		S = \int d^4x\sqrt{-g}\left[\dfrac{R}{2}-\dfrac{1}{f(\phi)}\left(\sqrt{1+f(\phi)\partial_\mu\phi\partial^{\mu}\phi}-1\right)-V(\phi)- \dfrac{1}{12}h^2(\phi)H_{\mu\nu\rho}H^{\mu\nu\rho}\right],
	\end{equation}
	where we have renamed the gauge kinetic function as $f(\phi)\to h(\phi)$ to avoid any misunderstanding. It turns out that the action \eqref{86} will become that of canonical scalar field considered in Refs. \cite{Ohashi:2013mka,Ohashi:2013qba,Ito:2015sxj} if the limit $f(\phi)\to 0$ is taken.
	
	Similar to the previous case, we will use the Bianchi type I metric shown in Eq. \eqref{3-2} to define the corresponding form of the action \eqref{86},
	\begin{equation}\label{88}
		S=\int d^4xe^{3\alpha}\left[3\dfrac{(\dot{\sigma}^2-\dot{\alpha}^2)}{N}-\dfrac{N}{f}\left(\sqrt{1-f\dfrac{\dot{\phi}^2}{N^2}}-1\right)-N V+\dfrac{h^2}{2N}e^{-4(\alpha+\sigma)}\dot{v}_B^2\right].
	\end{equation}
	As a result, the corresponding field equations of this model are given by
	\begin{align}\label{90}
		\dot{\alpha}^2& = \dot{\sigma}^2+\dfrac{\gamma^2}{3(\gamma+1)}\dot{\phi}^2+\dfrac{V}{3}+p_B^2\dfrac{e^{4\sigma-2\alpha}}{6h^2},\\
		\label{91}
		\ddot{\alpha }&=-\frac{3 \dot{\alpha }^2}{2}-\frac{3 \dot{\sigma }^2}{2}+\frac{V}{2}-\frac{\gamma  \dot{\phi }^2}{2 (\gamma +1)}+\frac{p_B^2 e^{4 \sigma -2 \alpha }}{12 h^2},\\\label{92}
		\ddot{\sigma }&=-3 \dot{\alpha } \dot{\sigma }-\frac{p_B^2 e^{4 \sigma -2 \alpha }}{3 h^2},\\
		\label{93}
		\ddot{\phi }&=-\frac{3 \dot{\alpha } \dot{\phi }}{\gamma ^2}-\frac{V_{\phi }}{\gamma ^3}-\frac{\gamma ^2 \dot{\phi }^4 f_{\phi }}{2 (\gamma +1)^2}-\frac{\gamma  \dot{\phi }^4 f_{\phi }}{(\gamma +1)^2}+\frac{p_B^2 e^{4 \sigma -2 \alpha } h_{\phi }}{\gamma ^3 h^3},
	\end{align}
	here we have used the following definition,
	\begin{equation}\label{key-17}
		f \equiv \dfrac{\gamma^2-1}{\gamma^2\dot{\phi}^2},
	\end{equation}
	as well as the solution of $v_B(t)$,
	\begin{equation}\label{89}
		\dot{v}_B = p_B h^{-2}\exp[\alpha+4\sigma].
	\end{equation}
	It is noted that $\gamma \equiv 1/\sqrt{1+ f(\phi)\partial_\mu \phi \partial^\mu \phi}$ is called the Lorentz factor characterizing the motion of the D3-brane \cite{Silverstein:2003hf,Copeland:2010jt}. The positivity of $f(\phi)$ implies that $\gamma > 1$.  The  model of canonical scalar field studied in Ref. \cite{Ito:2015sxj} will correspond to $\gamma=1$. 
	\subsection{Anisotropic power-law solution}
	With the above setting, to find exact anisotropic power-law solutions to the DBI model we continue to use the ansatz shown in Eq. \eqref{3-9} along with exponential functions such as \cite{Do:2011zz}
	\begin{align}
		V(\phi) &= V_0 \exp[\lambda\phi],\\
		f(\phi) &= f_0 \exp[\tau\phi],\\
		h(\phi) &= h_0 \exp[\rho\phi],
	\end{align}
	where $V_0$, $f_0$, $h_0$, $\lambda$, $\tau$, and $\rho$ are all constant. We also defined new variables \cite{Do:2011zz},
	\begin{align}
		u &= V_0 \exp[\lambda\phi_0],\\
		w &=p_B^2 h_0^{-2} \exp[-2\rho\phi_0],\\
		\kappa &= f_0 \exp[\tau\phi_0].
	\end{align}
	It turns out that the corresponding $\gamma$ is given by
	\begin{equation}\label{key-19}
		\gamma = \frac{1}{\sqrt{1-\kappa  \xi ^2 t^{-2+\xi  \tau }}}.
	\end{equation}
	Obviously, if $\tau\xi=2$ then $\gamma$ will become a constant $\gamma_0$ defined as
	\begin{equation}\label{key-20}
		\gamma = \gamma_0 = \dfrac{|\tau|}{\sqrt{\tau^2-4\kappa}}.
	\end{equation}
	Now, we are going to find out $\zeta$ and $\eta$ from the field equations because their values will determine whether the inflationary phase exists or not. Similar to the previous analysis, we will define a set of the algebraic equations from the field equations \eqref{90}, \eqref{91}, \eqref{92}, and \eqref{93} to be
	\begin{align}\label{104}
		\zeta ^2&=\frac{\gamma_0 ^2 \xi ^2}{3 \gamma_0 +3}+\eta ^2+\frac{u}{3}+\frac{w}{6},\\\label{105}
		-\zeta &=-\frac{\gamma_0  \xi ^2}{2 \gamma_0 +2}-\frac{3 \zeta ^2}{2}-\frac{3 \eta ^2}{2}+\frac{u}{2}+\frac{w}{12},\\\label{106}
		-\eta &=-3 \zeta  \eta -\frac{w}{3},\\\label{107}
		-\xi &=-\frac{3 \zeta  \xi }{\gamma_0 ^2}+\frac{2 \xi -2 \gamma_0  \xi }{\gamma_0 ^2+\gamma_0 }+\frac{\xi -\gamma_0  \xi }{\gamma_0 +1}+\frac{2 u}{\gamma_0 ^3 \xi }+\frac{\rho  w}{\gamma_0 ^3},
	\end{align}
	with the help of the corresponding constraints,
	\begin{align}
		\tau\xi&=2,\\
		\lambda\xi &=-2,\\	\label{109}
		-2\zeta+4\eta-2\xi\rho &= -2.
	\end{align}
	As a result, the relation between $\lambda$ and $\tau$ can be expressed as follows
	\begin{equation}\label{tau-lambda}
		\tau = -\lambda.
	\end{equation}
	Hence, the positivity of $\lambda$ will imply the negativity of $\tau$ and vice versa. 
	After some algebra, a set of non-trivial solutions is found to be
	\begin{align}\label{key-22}
		\zeta =&~\frac{4 \gamma_0 +\lambda ^2+5 \lambda  \rho +6 \rho ^2}{3 \lambda ^2+3 \lambda  \rho },\\
		\eta =&~\frac{4 \gamma _0+\lambda ^2+5 \lambda  \rho +6 \rho ^2}{2 \left(3 \lambda ^2+3 \lambda  \rho \right)}-\frac{\rho }{\lambda }-\frac{1}{2},\\
		u=&~\frac{2}{\left(\gamma _0+1\right) \lambda ^3 (\lambda +\rho )^2}\left[\gamma _0^2 \left(-\lambda ^2+4 \lambda  \rho +9 \rho ^2+4\right)\right.\nonumber\\
		&\left.+\gamma _0 \left(2 \lambda ^2 \rho ^2+\lambda ^2+7 \lambda  \rho ^3+8 \lambda  \rho +6 \rho ^4+11 \rho ^2\right)+4 \gamma _0^3+\rho ^2 \left(2 \lambda ^2+7 \lambda  \rho +6 \rho ^2\right)\right],\\
		w=&~\frac{4 \gamma _0 \lambda ^2-12 \gamma _0 \rho ^2-8 \gamma _0^2+4 \lambda ^3 \rho +14 \lambda ^2 \rho ^2+12 \lambda  \rho ^3}{\lambda ^2 (\lambda +\rho )^2}.
	\end{align}
	As a result, the scale factors for this model are given by
	\begin{align}\label{key-23}
		\zeta+\eta&=\frac{2 \gamma_0 +\rho  (\lambda +2 \rho )}{\lambda  (\lambda +\rho )},\\
		\zeta-2\eta&=\frac{2 \rho }{\lambda }+1.
	\end{align}
	Apparently, the positive scale factors require that both $\lambda$ and $\rho$ are positive definite. Therefore, $\tau$ should be negative according to Eq. \eqref{tau-lambda}. As a result,  the corresponding anisotropy parameter is given by
	\begin{equation}\label{119}
		\dfrac{\Sigma}{H} \equiv \dfrac{\dot{\sigma}}{\dot{\alpha}}=\dfrac{\eta}{\zeta}=\frac{2 \gamma_0 -\lambda  (\lambda +2 \rho )}{4 \gamma_0 +(\lambda +2 \rho ) (\lambda +3 \rho )}.
	\end{equation}
	It is straightforward to verify that these solutions will recover that defined in Ref. \cite{Ito:2015sxj} in the canonical limit $\gamma_0 \to 1$. 
	For the anisotropic power-law inflation with $\rho \gg \lambda$, the ratio $ \Sigma/H$ is indeed much smaller than one as expected. It appears that $\Sigma/H_{\text{DBI-two-form}} \simeq 0.0005 < \Sigma/H _{\text{canonical-two-form}}\simeq 0.0008$ provided $ \lambda=0.1$, $\rho=50$, and $\gamma_0=1.5$. This result indicates that the larger $\gamma_0$ is, the smaller ${\Sigma}/{H}$ is. Therefore, the  anisotropy parameter ${\Sigma}/{H}$ seems to be reduced in the context of non-canonical scalar field models, according to this result as well as that in the previous case. 
	\subsection{Stability analysis}
	The stability of the obtained anisotropic power-law solution of the DBI model will be investigated in this subsection. To do this task, we firstly introduce the corresponding dimensionless dynamical variables as
	\begin{equation}\label{key-24}
		x\equiv\dfrac{\dot{\sigma}}{\dot{\alpha}},\quad y \equiv\dfrac{\dot{\phi}}{\dot{\alpha}},\quad z \equiv p_B \dfrac{e^{-\alpha+2\sigma}}{h\dot{\alpha}}.
	\end{equation}
	Consequently, the field equations \eqref{90}, \eqref{91}, \eqref{92}, and \eqref{93} will be converted into the corresponding dynamical system of autonomous equations given by
	\begin{align}\label{123}
		\dfrac{dx}{d\alpha}&=3 x^3+\frac{x y^2}{2 \hat\gamma}+\frac{x z^2}{6}-3 x-\frac{z^2}{3},\\\label{124}
		\dfrac{dy}{d\alpha}&=3 \hat\gamma^2 \left[\hat\gamma \lambda  \left(x^2-1\right)-y\right]+3 x^2 y+\frac{y^3}{2 \hat\gamma}+y^2 \frac{2 \hat\gamma^2 (\lambda +\tau )-\hat\gamma \tau -\tau }{2 \left(\hat\gamma+1\right)}\nonumber\\
		&\quad+\frac{y z^2}{6}+\frac{1}{2} \hat\gamma^3 z^2 (\lambda +2 \rho ),\\\label{125}
		\dfrac{dz}{d\alpha}&=3 x^2 z+2 x z+\frac{y^2 z}{2 \hat\gamma}-\rho  y z+\frac{z^3}{6}-z,\\\label{126}
		\dfrac{d\hat\gamma}{d\alpha}&=-\frac{\left(\hat\gamma^2-1\right) }{6 \hat\gamma^2 y}\left[ y \hat\gamma\left(18 x^2-3 \tau  y+z^2\right)-6 \hat\gamma\dfrac{dy}{d\alpha}+3  y^3\right].
	\end{align}
	Here, we have defined an auxiliary variable as $\hat\gamma = 1/\gamma$ for the completeness of dynamical system \cite{Do:2011zz,Copeland:2010jt}.  Now, we are going to seek anisotropic fixed points with $x\neq 0$ to this dynamical system by solving a set of equations,
	\begin{equation}\label{key-25}
		\dfrac{dx}{d\alpha} = \dfrac{dy}{d\alpha}=\dfrac{dz}{d\alpha}= \dfrac{d\hat\gamma}{d\alpha}= 0.
	\end{equation}
	Combining both equations, ${dx}/{d\alpha} = 0$ and ${dz}/{d\alpha} = 0$, will yield an equation of $z^2$,
	\begin{equation}\label{128}
		z^2 = -3 x (2 x-\rho  y+2).
	\end{equation}
	On the other hand, the equation ${d\hat\gamma}/{d\alpha} = 0$ gives
	\begin{equation}\label{key-26}
		-\hat\gamma \left(18 x^2+z^2\right)-3 y^2+3 \hat\gamma\tau  y =0.
	\end{equation}
	Finally, the equation ${dz}/{d\alpha} = 0$ implies
	\begin{equation}\label{key-27}
		-6 \hat\gamma+18 \hat\gamma x^2+12 \hat\gamma x+3 y^2-6 \hat\gamma \rho  y+\hat\gamma z^2 =0. 
	\end{equation}
	Thanks to these useful results, an important expression for $y$ can be revealed to be
	\begin{equation}\label{131}
		y=\frac{2 (2 x-1)}{2 \rho -\tau }.
	\end{equation}
	Now, we are about to obtain an equation of $x$ by substituting Eq. \eqref{128} and Eq. \eqref{131} into the equation for ${dz}/{d\alpha} = 0$,
	\begin{equation}\label{key-28}
		\frac{(2 x-1) \left(-2 \hat{\gamma } \rho  \tau +\hat{\gamma } \tau ^2+6 \hat{\gamma } \rho ^2 x-5 \hat{\gamma } \rho  \tau  x+\hat{\gamma } \tau ^2 x+4 x-2\right)}{\hat{\gamma } (\tau -2 \rho )^2}=0.
	\end{equation}
	Solving this equation gives a non-trivial solution for $x$,
	\begin{align}
		x=\frac{\hat\gamma \tau  (2 \rho -\tau )+2}{\hat\gamma \left(6 \rho ^2-5 \rho  \tau +\tau ^2\right)+4}.
	\end{align}
	Very interestingly, the equation ${dy}/{d\alpha}=0$ can be rewritten as
	\begin{equation}\label{key-29}
		(\lambda+\tau)\dfrac{y}{6x^2}\left[-3 x^2-\frac{y^2}{\hat\gamma \left(\hat\gamma+1\right)}-\frac{z^2}{2}+3\right]=0,
	\end{equation}
	which clearly implies that $\tau=-\lambda$, or equivalently $\hat{\gamma} =\hat\gamma_0$ with $\hat\gamma_0 =1/\gamma_0$. This result is indeed consistent with the power-law solution found above. It is also consistent with Ref.
	\cite{Do:2011zz}, in which the DBI field is non-minimally coupled to the vector field. Consequently, a complete set of anisotropic fixed point solution could be represented in terms of $\lambda$, $\rho$, and $\hat\gamma_0$ as follows
	\begin{align}\label{135}
		x&=\frac{2-\hat\gamma_0 \lambda  (\lambda +2 \rho )}{\hat\gamma_0 \left(\lambda ^2+5 \lambda  \rho +6 \rho ^2\right)+4},\\
		y&=-\frac{6 \hat\gamma_0 (\lambda +\rho )}{\hat\gamma_0 \left(\lambda ^2+5 \lambda  \rho +6 \rho ^2\right)+4},\\\label{135c}
		z^2&=\frac{18 \left(\hat\gamma_0 \lambda ^2+2 \hat\gamma_0 \lambda  \rho -2\right) \left(2 \hat\gamma_0 \lambda  \rho +3 \hat\gamma_0 \rho ^2+2\right)}{\left(\hat\gamma_0 \lambda ^2+5 \hat\gamma_0 \lambda  \rho +6 \hat\gamma_0 \rho ^2+4\right)^2}.
	\end{align}

	It is clear that the positivity of $z^2$ will be ensured  if the following inequality,
		\begin{equation}\label{key-02}
			\hat\gamma_0 \lambda ^2+2 \hat\gamma_0 \lambda  \rho > 2,
		\end{equation}
		is satisfied. Furthermore, this anisotropic fixed point is exactly equivalent to the anisotropic power-law inflation solution found in the previous subsection. Therefore, we will examine the stability of the anisotropic fixed point solution, which will tell us the stability of the corresponding anisotropic power-law solution.
	
\begin{figure}[htp!]
		\centering
		\includegraphics[scale=0.6]{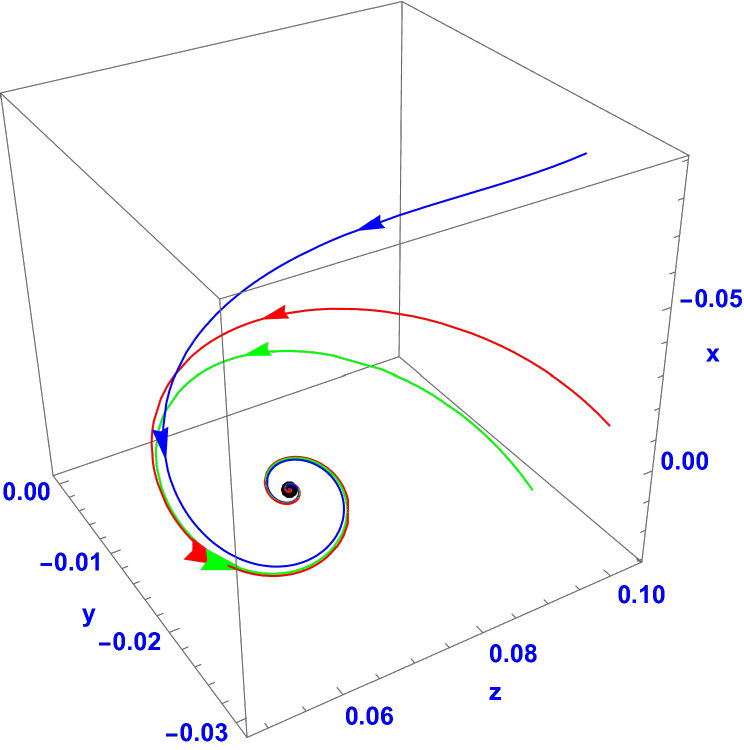}
		\caption{Attractive behavior of the anisotropic fixed point of the DBI case (displayed as a black point) for the parameters chosen as $\lambda = 0.1$, $\rho=50$, and $\gamma_0=1.5$.}
		\label{fig:numerical-DBI}
	\end{figure}

	To do this, we will first take approximated values for $x$, $y$, and $z^2$, during the inflationary phase,
	\begin{align}
		x&\simeq \frac{1-\hat\gamma_0 \lambda  \rho }{3 \hat\gamma_0 \rho ^2},\\
		y&\simeq -\dfrac{1}{\rho},\\
		z^2&\simeq \frac{3 \left(2\hat\gamma_0 \lambda  \rho+\hat\gamma_0\lambda^2 -2\right)}{2\hat\gamma_0 \rho ^2}.
	\end{align}
		Taking exponential perturbations, 
	\begin{align}
		\delta x &= A_x \exp[\omega\alpha],\\
		\delta y &= A_y \exp[\omega\alpha],\\
		\delta z &= A_z \exp[\omega\alpha],
	\end{align}
	we are able to define the following perturbed equations, which can be written in a matrix equation given by
	\begin{align}\label{}
		\mathcal{B}  \begin{pmatrix}
			A_x\\
			A_y\\
			A_{z}
		\end{pmatrix}&\equiv	
		\left(
		\begin{array}{ccc}
			-3-\omega & \frac{\hat{\gamma }_0 \lambda  \rho -1}{3 \hat{\gamma }_0^2 \rho ^3} & {-\frac{1}{\rho }\sqrt{\frac{2\lambda\hat\gamma_0 (\lambda +2 \rho )-4}{3\hat\gamma_0}}} \\
			-\frac{2 \left(\hat{\gamma }_0 \lambda  \rho -1\right) \left(\hat{\gamma }_0^3 \lambda  \rho -1\right)}{\hat{\gamma }_0 \rho ^3} & -3 \hat{\gamma }_0^2-\omega  & {6 \hat{\gamma }_0^3 \sqrt{\frac{\lambda \hat\gamma_0 (\lambda +2 \rho )-2}{6\hat{\gamma }_0}}} \\
			{\frac{ 6}{\rho } \sqrt{\frac{\lambda  \hat\gamma_0 (\lambda +2 \rho )-2}{6\hat\gamma_0}}}& {- \sqrt{\frac{3\lambda  \hat\gamma_0 (\lambda +2 \rho )-6}{2\hat\gamma_0}}} & {\frac{\lambda  (9 \lambda +10 \rho )}{12 \rho ^2}-\frac{1}{3 \hat{\gamma }_0 \rho ^2}-\omega}  \\
		\end{array}
		\right)
		\begin{pmatrix}
			A_x\\
			A_y\\
			A_{z}
		\end{pmatrix}=0.
	\end{align}
	Consequently, the corresponding equation for $\omega$ is defined as
	\begin{align}\label{148}
		b_3\omega ^3+	b_2 \omega ^2 +b_1 \omega +  b_0=0,
	\end{align}
	where 
	\begin{align}
		b_3&=1>0,\\
		b_2 &\simeq 3 \left(\hat{\gamma_0}^2+1 \right)>0,\\
		b_1 &\simeq 3\hat\gamma_0^2 \left(2\lambda\rho\hat\gamma_0+1 \right)>0,\\
		b_0 &\simeq {9\hat\gamma_0^2 \left(2\lambda\rho\hat\gamma_0+\lambda^2\hat\gamma_0-2 \right)} >0.
	\end{align}
	It is noted that the positivity of the last coefficient $b_0$ is ensured by the positivity of $z^2$. It is crystal that Eq. \eqref{148} with all positive coefficients will always admit negative roots of $\omega$. Thereby, the anisotropic fixed point of the DBI case is indeed stable during the inflationary phase. Consistently, the numerical result displayed in Fig. \ref{fig:numerical-DBI} points out that this fixed point is attractive. It is safe to conclude that the anisotropic power-law of the DBI case is stable and disfavors the cosmic no-hair conjecture. 
	\section{Comparisons of $|\Sigma/H|$ for different models} \label{sec5}
	\subsection{Two-form field models}
\begin{figure}[htp!]
		\centering
		\begin{subfigure}[b]{0.45\textwidth}
			\includegraphics[scale=1]{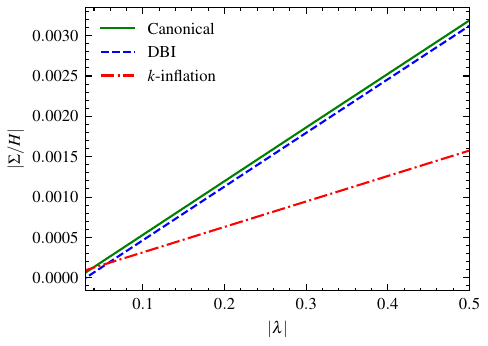}
			\caption{$|\Sigma/H|$ varies with $|\lambda| $}
			\label{fig:lambda-ani}
		\end{subfigure}
		\qquad
		\begin{subfigure}[b]{0.45\textwidth}
			\includegraphics[scale=1]{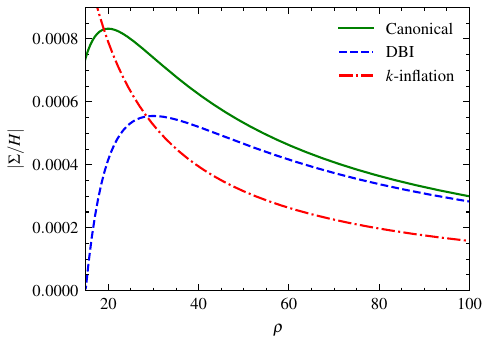}
			\caption{$|\Sigma/H|$ varies with $\rho $}
			\label{fig:rho-ani}
		\end{subfigure}
		\caption{(Left) The anisotropy parameter $|\Sigma/H|$ as a function of $|\lambda|$ for the fixed  values, $\rho = 50$ and $\gamma_0 =1.5$. (Right) The anisotropy parameter $|\Sigma/H|$ as a function of $\rho$ for the fixed values, $|\lambda| = 0.1$ and $\gamma_0 =1.5$.}
		\label{fig:three graphs}
	\end{figure}

	 For heuristic reasons, we would like to compare the anisotropy parameter, $|\Sigma/H|$, for three different anisotropic power-law inflation models of two-form field. As a result, the expression of $|\Sigma/H|$ for the canonical scalar field can be found in Eq. (48) of Ref. \cite{Ito:2015sxj}. According to Fig. \ref{fig:three graphs}, it appears that $|\Sigma/H|$ varies at different rates for the canonical, {\it k}-inflation, and DBI scalar fields. Specifically, Fig. \ref{fig:lambda-ani} contains three curves of $|\Sigma/H|$ for the fixed values $\rho = 50$ and $\gamma_0=1.5$, along with $|\lambda|$ running from ${0.03}$ to $0.5$. On the other hand, Fig. \ref{fig:rho-ani} is plotted for the fixed values $|\lambda| = 0.1$ and $\gamma_0=1.5$, along with $ \rho$ varying from ${15}$ to $100$. {Note that the ranges for $|\lambda|$ and $\rho$ have been chosen to accommodate the stability conditions for the canonical model \cite{Ito:2015sxj} ($\lambda> \sqrt{\rho ^2+2}-\rho\simeq 0.02$ for a fixed value of $\rho=50$ and $\rho > {1}/{\lambda }-{\lambda }/{2}= 9.95$ for a fixed value of $\lambda=0.1$) as well as the DBI model \eqref{key-02} ($\lambda> {\sqrt{\hat{\gamma }_0^2 \rho ^2+2 \hat{\gamma }_0}}/{\hat{\gamma }_0}-\rho\simeq 0.03$ for fixed values $\rho=50$ and $\gamma_0 = 1.5$ and $\rho > {1}/({\hat{\gamma }_0 \lambda })-{\lambda }/{2}= 14.95$ for fixed values  $\lambda=0.1$ and $\gamma_0 = 1.5$). Also note that, these ranges are automatically consistent with the stability of the {\it k}-inflation} model. It is crystal, according to these two figures, that the values of $|\Sigma/H|$  for non-canonical scalar fields, i.e., the {\it k}-inflation and DBI ones, are always smaller than that of the canonical scalar one during the inflationary phase with $\rho  \gg |\lambda|$. More interestingly, the value of $|\Sigma/H|$ for the {\it k}-inflation case turns out to be smallest among three types of scalar fields.  This result would be useful to investigate the observational constraints of anisotropic inflation due to the two-form field. 
\subsection{One-form field vs. two-form field}
\begin{figure}[htp!]
		\centering
		\begin{subfigure}[b]{0.45\textwidth}
			\includegraphics[scale=1]{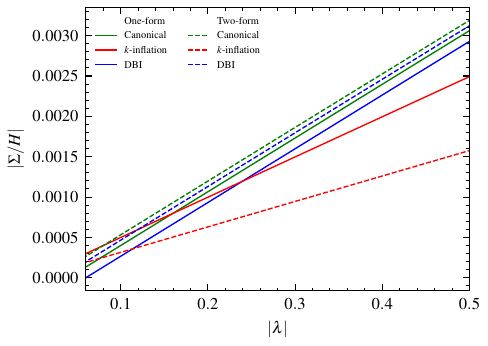}
			\caption{$|\Sigma/H|$ varies with $|\lambda| $}
			\label{fig:lambda-ani-12}
		\end{subfigure}
		\qquad
		\begin{subfigure}[b]{0.45\textwidth}
			\includegraphics[scale=1]{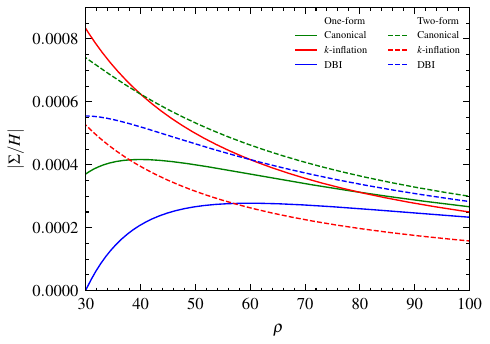}
			\caption{$|\Sigma/H|$ varies with $\rho $ }
			\label{fig:rho-ani-12}
		\end{subfigure}
		\caption{Comparisons among six anisotropy parameters $|\Sigma/H|$ derived in two types of anisotropic  power-law inflation, one is due to the one-form field and the other is due to the two-form field, for three different kinds of scalar fields. The left figure is plotted for the fixed values, $\rho=50$ and $\gamma_0 =1.5$, while the right figure is plotted for the fixed values, $|\lambda|=0.1$ and $\gamma_0=1.5$.}
		\label{fig:three graphs-12}
	\end{figure}

One might ask about more comparisons among six anisotropy parameters $|\Sigma/H|$ derived in two types of anisotropic power-law inflation, one is due to the one-form field (a.k.a. vector field) and the other is due to the two-form field, for three different kinds of scalar fields. In this subsection,  therefore, we would like to make such comparisons. First, we should note that the derivations of $|\Sigma/H|$ for one-form anisotropic inflation have been shown in Eq. (2.34) of Ref. \cite{Kanno:2010nr}, Eq. (3.34) of Ref. \cite{Do:2011zz}, and Eq. (3.21) of Ref. \cite{Do:2020hjf} for the canonical, DBI, and {\it k}-inflation scalar fields, respectively. {In particular, Fig. \ref{fig:lambda-ani-12} contains six curves of $|\Sigma/H|$ for the fixed values $\rho = 50$ and $\gamma_0=1.5$, along with {$|\lambda|\in [0.06,0.5]$}. On the other hand, Fig. \ref{fig:rho-ani-12} is plotted for the fixed values $|\lambda| = 0.1$ and $\gamma_0=1.5$, along with $ \rho \in [30,100]$}. {Once again, the value ranges of $|\lambda|$ and $\rho$ have been chosen to accommodate the stability conditions for all six models we have been interested in}. According to two Figs. \ref{fig:lambda-ani-12} and \ref{fig:rho-ani-12}, we observe that the $|\Sigma/H|$ of {\it k}-inflation two-form anisotropic inflation tends to be one of the smallest ones among that of six models when $\rho$ increases or $|\lambda|$ decreases during the inflationary phase. More interestingly, it turns out that the values of $|\Sigma/H|$ of canonical and DBI two-form anisotropic inflation are always larger than that of canonical and DBI one-form anisotropic inflation, respectively. In contrast, $|\Sigma/H|$ of {\it k}-inflation two-form anisotropic inflation is always smaller than that of {\it k}-inflation one-form anisotropic inflation. All these remarkable points would be useful when judging which anisotropic inflation model is cosmologically viable in the light of observational data. 
\section{Tensor-to-scalar ratio}\label{sec6}
In connection with the Planck 2018 data \cite{Planck:2018vyg,Planck:2018jri} as well as  future detections like the CMB-S4 project \cite{CMB-S4:2020lpa}, we would like to investigate in this section the corresponding tensor-to-scalar ratio of the present non-canonical anisotropic inflationary models, following our previous works \cite{Do:2016ofi,Do:2020hjf,Do:2020ler} for the one-form field as well as other works in Refs. \cite{Watanabe:2010fh,Dulaney:2010sq,Gumrukcuoglu:2010yc,Watanabe:2010bu,Bartolo:2012sd,Chen:2014eua} for the one-form field and in Refs. \cite{Ohashi:2013mka,Ohashi:2013qba} for the two-form field.

As discussed earlier, the anisotropy deviation $\sigma$ should be much smaller than the isotropy parameter $\alpha$ in order to be consistent with the observations of WMAP and Planck. Therefore, it is reasonable to regard the background metric as the spatially flat Friedmann-Lemaitre-Robertson-Walker (FLRW) metric, rather than the Bianchi type I metric, for simplicity as done in Refs. \cite{Do:2016ofi,Do:2020hjf,Do:2020ler,Watanabe:2010fh,Dulaney:2010sq,Gumrukcuoglu:2010yc,Watanabe:2010bu,Bartolo:2012sd,Chen:2014eua} for the one-form field as well as Refs. \cite{Ohashi:2013mka,Ohashi:2013qba} for the two-form field. 

It is important to mention that when the statistical isotropy of CMB is broken down the scalar power spectrum will be modified, according to Ref. \cite{Ackerman:2007nb}, to be
\begin{equation} \label{ACW}
	{\cal P}^{\tilde{\zeta}}_{k,\text{ani}} = {\cal P}^{\tilde{\zeta}(0)}_k\left(1+g_\ast \cos^2\theta \right),
\end{equation}
where $g_\ast$ is a non-trivial constant characterizing a deviation from the spatial isotropy and is expected to be smaller than one, i.e., $|g_\ast| < 1$. In addition, ${\cal P}^{\tilde{\zeta}(0)}_k $ is nothing but the isotropic scalar power spectrum, while $\theta$ is the angle between the wave number ${\bf k}$ and the  privileged direction $\bf{V}$ close to the ecliptic poles \cite{Ackerman:2007nb}. On the observation side, it is worth noting that small values of $g_\ast$ have been identified via several analyses in Refs. \cite{Groeneboom:2009cb,Kim:2013gka,Ramazanov:2013wea,Ramazanov:2016gjl,Sugiyama:2017ggb}. On the theoretical side,  the analytical formula of $g_\ast$ has been derived within the context of the anisotropic inflation of one-form field in several published papers, e.g., in Refs. \cite{Watanabe:2010fh,Ohashi:2013qba} for the canonical scalar field and in Refs. \cite{Do:2016ofi,Do:2020hjf,Do:2020ler} for the non-canonical scalar field. More interestingly, the analytical formula of $g_\ast$ for the anisotropic inflation of two-form field has been determined in Refs. \cite{Ohashi:2013mka,Ohashi:2013qba}. However, this formula is limited only to the canonical scalar field. We will therefore derive in this section a more general formula of $g_\ast$ for non-canonical scalar fields.
\subsection{Scalar perturbations}
Following our previous works done in Refs. \cite{Do:2016ofi,Do:2020ler} as well as the seminal paper \cite{Garriga:1999vw}, the metric of scalar perturbations is given by
\begin{equation}\label{}
	ds^2 = a^2(\tilde{\eta})\left[-(1+\Phi)d\tilde{\eta}^2+(1+2\Psi)\delta_{ij}dx^idx^j\right],
\end{equation}
with $\tilde{\eta}\equiv \int a^{-1}dt$ is a conformal time.  In order to compute the scalar power spectrum, we adopt the spatially flat gauge with ${\Psi} = 0$, which has been used in our previous paper \cite{Do:2020ler}
\begin{equation}\label{}
	\tilde{\zeta} = -\dfrac{H}{\dot{\phi}}\delta\phi,
\end{equation}
here $H$ is the Hubble parameter. It is noted that this gauge has also been used in many previous papers on anisotropic inflation, e.g., see Refs. \cite{Ohashi:2013mka,Ohashi:2013qba,Chen:2014eua}. It is also noted that we have used modified notations such as $\tilde{\eta}$ and $\tilde{\zeta}$, since $\zeta$ and $\eta$ have been used for the power-law solution in the previous sections.

In the absence of non-minimal coupling between the scalar and two-form fields, the isotropic scalar power spectrum ${\cal P}^{\tilde{\zeta}(0)}_k$ can be obtained \cite{Do:2020hjf,Do:2020ler}
\begin{equation}\label{}
	{\cal P}^{\tilde{\zeta}(0)}_k= {\cal P}^{\tilde{\zeta}(0)}_{k,\text{nc}}=\left.\dfrac{1}{8\pi^2 M_p^2}\dfrac{H^2}{c_s\epsilon}\right|_{c_s^\ast k_\ast = a_\ast H_\ast},
\end{equation}
which is identical to that firstly obtained in Ref. \cite{Garriga:1999vw}. Here, 
the notation `$\ast$' implies the pivot scale (a.k.a. horizon-exit scale), where the spacetime of universe can be approximated as the de Sitter one with  $a_\ast \simeq -(H_\ast \tilde{\eta}_\ast)^{-1}$, while the superscript $(0)$ stands for the de Sitter background.  In addition, $\epsilon$ is the slow-roll parameter defined as $\epsilon \equiv -\dot{H}/H^2$, while $c_s$ is the speed of sound, whose definition is given by \cite{Garriga:1999vw}
\begin{equation}
	c_s^2 \equiv \dfrac{\partial_X p}{\partial_X\rho}=\frac{\partial_X P(\phi,X)}{\partial_X[2X \partial_XP(\phi,X)-P(\phi,X)]},
	\end{equation}
where $p$ and $\rho$ the pressure and energy density parameters, respectively.

In the presence of non-minimal coupling between the non-canonical scalar field and the two-form field, the full scalar power spectrum turns can be calculated, following our previous works \cite{Do:2016ofi,Do:2020ler} as well as Refs. \cite{Ohashi:2013mka,Ohashi:2013qba}, to be
\begin{equation}\label{}
	{\cal P}^{\tilde{\zeta}}_{k,\text{nc}}   = {\cal P}^{\tilde{\zeta}(0)}_k+\delta {\cal P}^{\tilde{\zeta}}_k ,
\end{equation}
where the correction term $\delta{\cal P}^{\tilde{\zeta}}_k$ is given by (see the Appendix \ref{appA} for derivations)
\begin{align}\label{}
	\delta {\cal P}^{\tilde{\zeta}}_k = \dfrac{E_{yz}^2c_s^4 N_{c_s k}^2\cos^2\theta}{4 \pi^2 \epsilon^2 M_p^4 }.
\end{align}
Here, $E_{yz} \equiv h/a^3 H_{0yz} = h/a^3 B'_{yz}$ is the background value of the two-form field \cite{Ohashi:2013mka,Ohashi:2013qba}, $a\equiv e^\alpha = -(\tilde{\eta} H)^{-1}$ is a scale factor of  the background de Sitter spacetime, $'$ represents a derivative with respect to the conformal time $\tilde{\eta}$, i.e.,  $B'_{yz}=d B_{yz}/d\tilde \eta$, and $h$ is nothing but the gauge kinetic function. It is noted that the key element to derive the above formula is the standard Bunch-Davies (BD) vacuum state for the non-canonical scalar field, whose definition has been shown in Ref.  \cite{Chen:2006nt} as
\begin{equation}\label{BD}
	\tilde{\zeta}^{(0)}_k(\tilde{\eta}) = \dfrac{H}{2\sqrt{c_s \epsilon}M_p k^{3/2}}(1+ic_s k\tilde{\eta}) e^{-ic_sk\tilde{\eta}}.
\end{equation}
This BD vacuum state has also been used in our previous papers \cite{Do:2016ofi,Do:2020ler} to compute the CMB imprints of the non-canonical anisotropic inflation based on the one-form field. 
 As a result, the full scalar power spectrum can be expressed as
\begin{align}
		{\cal P}^{\tilde{\zeta}}_{k,\text{nc}}   =  {\cal P}_{k,\text{nc}}^{\tilde{\zeta}(0)} \left(1+\dfrac{2c_s^5 E_{yz}^2 N_{c_sk}^2}{\epsilon H^2 M_p^2}\cos^2\theta\right),
\end{align}
with $N_{c_s k }\simeq 60 $ is the e-fold number.  Therefore, the corresponding $g_\ast$ for non-canonical scalar fields can be defined as
\begin{equation}\label{def-g}
	g_\ast = c_s^5\dfrac{2 E_{yz}^2 N_{c_sk}^2}{\epsilon H^2 M_p^2}  =  c_s^5 g^0_\ast>0,
\end{equation}
where
\begin{equation} \label{def-g-0}
	g_\ast^0 = \dfrac{2 E_{yz}^2 N_{c_sk}^2}{\epsilon H^2 M_p^2} >0,
\end{equation}
for the canonical scalar field \cite{Ohashi:2013mka,Ohashi:2013qba}.  Subsequently, the scalar spectral index can be shown as
\begin{align}\label{6.16}
	n_s - 1 \equiv \left.\dfrac{d\ln {\cal P}^{\tilde{\zeta}}_{k,\text{nc}}}{d\ln k}\right|_{c_s^\ast k_\ast= a_\ast H_\ast}\simeq -2\epsilon-\bar{\eta}-s-\left(\dfrac{2}{N_{c_s k}}-5s\right)\dfrac{g_\ast \cos^2\theta}{1+g_\ast \cos^2\theta },
\end{align}
where $\bar{\eta} \equiv \dot{\epsilon}/(\epsilon H)$ and $s\equiv \dot{c}_s/(c_s H)$ \cite{Garriga:1999vw,Baumann:2006cd}. Given that the average value of $\cos^2\theta$ is $\langle\cos^2\theta\rangle = 1/3$ \cite{Ohashi:2013mka,Ohashi:2013qba}, Eq. \eqref{6.16} now becomes
\begin{equation}\label{}
	n_s-1 \simeq-2\epsilon - \bar{\eta}-s - \left(\dfrac{2}{N_{c_s k}}-5s\right)\dfrac{g_\ast}{3+g_\ast}.
\end{equation}
It is important to note that $g_\ast^0$ and $g_\ast$ have been shown to be negative definite for the one-form field \cite{Do:2016ofi,Do:2020hjf,Do:2020ler,Watanabe:2010fh,Ohashi:2013qba}, while they turn out to be positive definite for the two-form field, according to Eqs. \eqref{def-g} and \eqref{def-g-0}. This difference is a very interesting point, which could be useful to distinguish two types of anisotropic inflation, one is due to the one-form field and the other is due to the two-form field, in the light of observational constraints \cite{Groeneboom:2009cb,Kim:2013gka,Ramazanov:2013wea,Ramazanov:2016gjl,Sugiyama:2017ggb}.
\subsection{Tensor perturbations}
In this subsection, we consider tensor perturbations for a model of non-canonical scalar fields non-minimally coupled to the two-form field. It is important to note that tensor perturbations for models of a canonical scalar field non-minimally coupled to either the one-form or two-form field have been investigated in the literature, e.g., see Refs. \cite{Ohashi:2013qba,Ohashi:2013mka}.  Additionally, tensor perturbations for a model of non-canonical scalar fields non-minimally coupled to the one-form field have been presented in our previous papers \cite{Do:2016ofi,Do:2020ler}. According to these papers, the metric of tensor perturbations can be expressed as \begin{align}
	g_{\mu\nu} = a^2(\tilde{\eta})\left[-d\tilde{\eta}^2+(\delta_{ij}+h_{ij})dx^idx^j\right],
\end{align}
with $h_{ij}$ is the traceless ($\delta^{ij}h_{ij} = 0$) and transverse ($\partial^ih_{ij} = 0$) tensor perturbations satisfying the condition $|h_{ij}|\ll 1$. As a result, $h_{ij}$ has two degrees of freedom,  $h_+$ and $h_\times$, which are denoted as polarizations. Since the gravitational sector is not affected by the scalar field as well as the non-minimal coupling between the non-canonical scalar and two-form fields \cite{Ohashi:2013qba,Ohashi:2013mka}, then the tensor power spectrum for anisotropic inflation remains identical to that for isotropic inflation, which can be found in Ref. \cite{Garriga:1999vw}, i.e.,
\begin{equation}
	{\cal P}^h_{k,\text{nc}} = {\cal P}^{h(0)}_{k,\text{nc}} = 16 c_s\epsilon{\cal P}^{\tilde{\zeta}(0)}_{k,\text{nc}}.
\end{equation}
Consequently, the tensor spectral index reads
\begin{align}
	n_t \equiv \left. \dfrac{d\ln {\cal P}^h_{k,\text{nc}}}{d\ln k}\right|_{k_\ast=a_\ast H_\ast}\simeq -2\epsilon.
\end{align}
\subsection{Full tensor-to-scalar ratio}
Based on the results derived above, we now end up with the full tensor-to-scalar ratio for the two-form case as follows
\begin{align}
	r_{\text{nc}} \equiv \dfrac{{\cal P}^{h}_{k,\text{nc}}}{{\cal P}^{\tilde{\zeta}}_{k,\text{nc}}} = \dfrac{16c_s\epsilon}{1+c_s^5 g^0_\ast \cos^2\theta},
\end{align}
which can be reduced to
\begin{align}
	r_{\text{nc}} = 16c_s\epsilon\dfrac{3}{3+c_s^5g_\ast^0},
\end{align}
if we take $\langle\cos^2\theta\rangle = 1/3$ \cite{Ohashi:2013mka,Ohashi:2013qba}. Apparently,  this  tensor-to-scalar ratio will recover that derived in Ref. \cite{Ohashi:2013qba},
\begin{align}
	r_{\text{nc}} \to r =  16\epsilon{{\dfrac{3}{3+g_\ast^0}}},
\end{align}
in the canonical limit $c_s \to 1$. 
\subsection{Application to the anisotropic power-law solutions}
The tensor-to-scalar ratio is perhaps the most significant parameter of any inflationary model, by which we can judge how viable it is in the light of Planck 2018 data \cite{Planck:2018vyg,Planck:2018jri} or future detections like the CMB-S4 project \cite{CMB-S4:2020lpa}. Therefore, we will consider in this subsection whether the obtained anisotropic power-law inflationary solutions are consistent with the Planck 2018 data. It is noted that the results derived in the previous subsections are general, which could be valid for many types of non-canonical scalar field.

We first consider the {\it k}-inflation model, which was firstly mentioned  in the one-form case \cite{Do:2020hjf} and has been investigated in Sect. \ref{sec3}, with the corresponding $P(\phi,X)$ is given by
\begin{equation}\label{}
	P(\phi,X) = k_0 X+l_0 e^{\lambda\phi}X^2.
\end{equation}
As a result, the corresponding speed of sound turns out to be
\begin{equation}
	c_s^2 = \dfrac{k_0+2l_0e^{\lambda\phi}X}{k_0+6l_0e^{\lambda\phi}X}.
\end{equation}
Furthermore, $c_s^2$ becomes, for the anisotropic power-law inflationary solution derived in Sect. \ref{sec3}, as
\begin{align}
	c_s^2 = \dfrac{k_0+u \xi^2}{k_0+3 u\xi^2} \simeq -\dfrac{5\lambda}{228\rho}\simeq \dfrac{5}{114\zeta}\ll 1
\end{align}
along with the corresponding slow-roll parameter given by
\begin{equation}
	\epsilon \equiv -\dfrac{\dot{H}}{H^2}= \dfrac{1}{\zeta}\simeq \dfrac{114}{5}c_s^2\ll 1.
\end{equation}

Next, we consider the DBI inflation model with the corresponding form of $P(\phi,X)$ given by
\begin{equation}
	P(\phi,X) = -\dfrac{1}{f(\phi)}\left(\sqrt{1-2Xf(\phi)}-1\right)-V(\phi).
\end{equation}
It turns out that the corresponding speed of sound is defined as
\begin{equation}
	c_s^2 = \dfrac{1}{\gamma^2}.
\end{equation}
 It is apparent that $\gamma =\gamma_0$ for the power-law inflation as pointed out in Sect. \ref{sec4}, i.e., $c_s^2=\gamma_0^{-2}$.
In addition, the corresponding slow-roll parameter reads
\begin{equation}\label{key}
	\epsilon \equiv -\dfrac{\dot{H}}{H^2}=\dfrac{1}{\zeta}\simeq\dfrac{\lambda}{2\rho}.
\end{equation}
It is clear that $\epsilon$ will not be expressed in terms of $c_s$ in this DBI case, unlike the {\it k}-inflation case. 

Due to the fact that  $\bar{\eta} = s =0$ for the power-law solutions, the corresponding scalar spectral index for both models reads
\begin{equation}
	n_s-1\simeq -2\epsilon -\frac{2 c_s^5 g^0_\ast }{N_{c_sk}\left(3+c_s^5g^0_\ast \right) },
\end{equation}
where $N_{c_sk} \simeq 60$ will be chosen as usual.

\begin{figure}[hbtp]
	\begin{center}
	\includegraphics[scale=0.5]{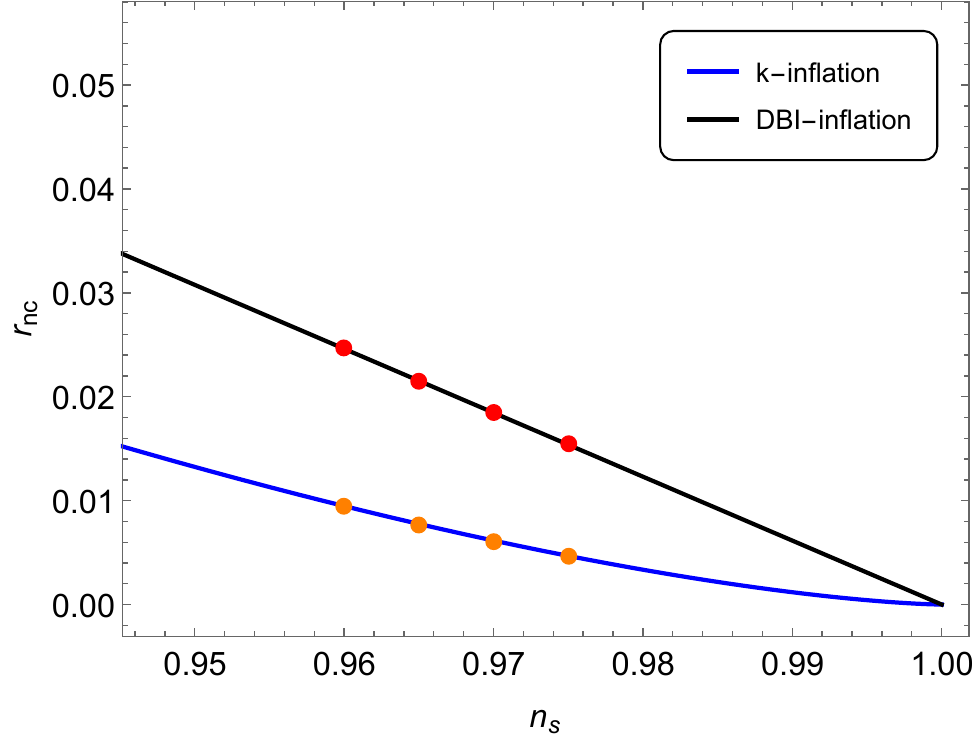}
	\caption{The $n_s -r_{\rm nc}$ diagram of anisotropic power-law inflationary solution of {\it k}-inflation two-form field model (lower curve) and that of DBI two-form field model (upper curve) are plotted all together for comparisons. Four colored points are displayed to indicate a prediction region, which overlaps with the most confidence level region for $n_s$ of the Planck 2018 data, in which $0.96\leq n_s \leq 0.975$.}
	\label{fig:tts-k-dbi}
	\end{center}
\end{figure}

\begin{figure}[hbtp]
	\begin{center}
	\includegraphics[scale=0.5]{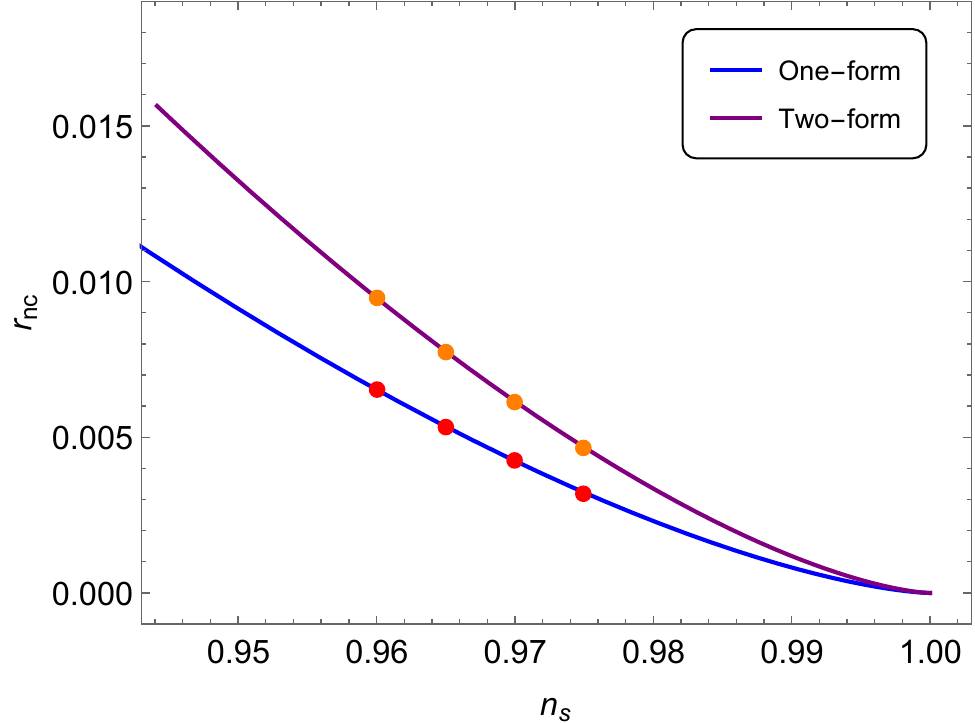}
	\caption{The $n_s -r_{\rm nc}$ diagram of anisotropic power-law inflationary solution of {\it k}-inflation one-form field model (lower curve) vs. that of {\it k}-inflation two-form field model (upper curve). Four colored points are displayed to indicate a prediction region, which overlaps with the most confidence level region for $n_s$ of the Planck 2018 data, in which $0.96\leq n_s \leq 0.975$.}
	\label{fig:tts-k-one-two-form}
	\end{center}
\end{figure}

In order to compare with the Planck 2018 data \cite{Planck:2018vyg,Planck:2018jri}, we are going to plot the $n_s -r_{\rm nc}$ diagram. First, we will choose $g^0_\ast = +0.03$, according to the analytical formula shown in Eq. \eqref{def-g-0} as well as the recent observational constraints of $g_\ast^0$ published in Refs. \cite{Kim:2013gka,Ramazanov:2013wea,Ramazanov:2016gjl,Sugiyama:2017ggb} for both {\it k}-inflation and DBI models.  In addition, we consider a range for the speed of sound as  $0< c_s \leq 0.035$ for the {\it k}-inflation model. For the DBI inflation model, we choose a range of the ratio $\lambda/\rho $  as $0<\lambda/\rho\leq0.055$ along with the fixed value for speed of sound is $c_s \simeq 0.077$ (equivalent to $\gamma=\gamma_0 = 13$). According to Fig. \ref{fig:tts-k-dbi}, it is obvious that the {\it k}-inflation model turns out to be more viable, in the light of the Planck 2018 data, than the DBI one. Indeed, the anisotropic power-law inflationary of the {\it k}-inflation model is highly consistent with the Planck 2018 data \cite{Planck:2018vyg,Planck:2018jri}, similar to the solution found in Ref. \cite{Do:2020hjf} for the one-form field.  More interestingly, it appears that the tensor-to-scalar ratio of {\it k}-inflation one-form field model \cite{Do:2020hjf} turns out to be smaller than that of {\it k}-inflation two-form field model (see Fig. \ref{fig:tts-k-one-two-form} for the detailed comparison). It should be noted that we have used a more precise formula, $c_s^2 \simeq -\lambda/(48 \rho)$, instead of an approximated formula shown in Eq. (5.15) of Ref. \cite{Do:2020hjf} in order to plot the $n_s -r_{\rm nc}$ diagram of  {\it k}-inflation one-form field model as displayed in Fig. \ref{fig:tts-k-one-two-form}. Remarkably, the tensor-to-scalar ratios of both {\it k}-inflation one-form and two-form models seem to be relevant to a  forecast of the CMB-S4 project \cite{CMB-S4:2020lpa}. 

	\section{Conclusions} \label{final}
	We have studied the anisotropic power-law inflation in the presence of the two-form field non-minimally coupled to non-canonical scalar fields. For this purpose, we have focused on examining two typical non-canonical forms of scalar field, which come from the  {\it k}-inflation \cite{Do:2020hjf,Armendariz-Picon:1999hyi,Garriga:1999vw} and DBI inflation \cite{Do:2011zz,Ohashi:2013pca,Holland:2017cza,Nguyen:2021emx,Silverstein:2003hf,Copeland:2010jt}. As a result, the anisotropic inflationary solutions derived in these models become stable and attractive as confirmed by the stability analysis based on the dynamical system method. Our present work together with the previous ones by other people in Refs. \cite{Ohashi:2013qba,Ohashi:2013mka,Ito:2015sxj,Do:2018zac,Almeida:2019xzt} demonstrate that the cosmic no-hair conjecture proposed by Hawking et al. \cite{Gibbons:1977mu} is extensively violated due to the existence of non-minimal coupling between the scalar and two-form fields such as $h^{2}( \phi)H_{\mu \nu\rho}H^{\mu\nu\rho}$. More interestingly, we have pointed out, by taking some simple comparisons, that the non-canonical property of scalar field seems to reduce the magnitude of anisotropy parameter $|\Sigma/H|$. This interesting point would be useful when investigating observational constraints of the anisotropic inflation based on two-form field. Additionally, we have made comparisons among  six anisotropy parameters $|\Sigma/H|$ derived in two types of anisotropic power-law inflation, one is due to the one-form field and the other is due to the two-form field, for three different kinds of scalar fields.  As a result, the $|\Sigma/H|$ of {\it k}-inflation two-form inflation tends to be one of the smallest ones among six anisotropy parameters. This additional interesting point would also be useful when judging which anisotropic inflation model is cosmologically viable in the light of observational data.  In connection with the Planck 2018 data \cite{Planck:2018vyg,Planck:2018jri} and future detections like the CMB-S4 project \cite{CMB-S4:2020lpa}, we have investigated the tensor-to-scalar ratio of the obtained anisotropic power-law solutions. It turns out that the tensor-to-scalar ratio of the anisotropic power-law solution of {\it k}-inflation model is more consistent with the Planck 2018 data than that of DBI model. More interestingly, additional analysis has shown that the tensor-to-scalar ratio of {\it k}-inflation one-form field model is more viable than that of {\it k}-inflation two-form field model in the light of Planck 2018 data \cite{Planck:2018vyg,Planck:2018jri}. Furthermore, these tensor-to-scalar ratios seem to be relevant to a forecast of the CMB-S4 project \cite{CMB-S4:2020lpa}.  We hope that our study would contribute a new perspective on counterexamples to the cosmic no-hair theorem. It should be noted that the research on the CMB imprints of anisotropic inflation model, in which the canonical scalar field is non-minimally coupled to the two-form field,  has been investigated in Refs. \cite{Ohashi:2013mka,Ohashi:2013qba}. Our next step would therefore be a study of the effect of non-canonical scalar fields on the CMB imprints of two-form anisotropic inflation like what we have done in Ref. \cite{Do:2020ler}. Besides, we will continue to examine whether anisotropic inflation could be caused by other non-minimal couplings between scalar and other fields such as the Yang-Mills one \cite{Gomez:2021jbo}. We leave these issues for our future works.
	\section*{Acknowledgments}
	We would like to thank the referee very much for useful comments and suggestions.
	W.F.K. is supported in part by the NSTC of Taiwan under Contract No. NSTC 110-2112-M-A49-007. We would like to thank Prof. Sergey Odintsov, Prof. Tanmoy Paul, Prof. Eoin O Colgain, and Dr. Gabriel Gomez very much for their useful comments and suggestions. 
	\appendix
	\section{The second order correction to the full scalar power spectrum}\label{appA}
	It turns out that the  full power spectrum can be calculated as follows \cite{Ohashi:2013mka,Ohashi:2013qba}
	\begin{align}
		{\cal P}_k^{\tilde{\zeta}}(\tilde{\eta},k) = &~ \langle0|\hat{\tilde{\zeta}}_k(\tilde{\eta})\hat{\tilde{\zeta}}_{k'}(\tilde{\eta})|0\rangle\nonumber\\
		 \simeq &~ \langle0|\hat{\tilde{\zeta}}_k^{(0)}(\tilde{\eta})\hat{\tilde{\zeta}}_{k'}^{(0)}(\tilde{\eta})|0\rangle - i\int_{\tilde{\eta}_{min}}^{\tilde{\eta}} \langle0|[H_{\tilde{\zeta}}(\tilde{\eta}_1),\hat{\tilde{\zeta}}^{(0)}_k(\tilde{\eta})\hat{\tilde{\zeta}}^{(0)}_{k'}(\tilde{\eta})]|0\rangle d\tilde{\eta}_1\nonumber\\
		&+\int_{\tilde{\eta}_{min,1}}^{\tilde{\eta}}d\tilde{\eta}_1\int_{\tilde{\eta}_{min,2}}^{\tilde{\eta}_1} d\tilde{\eta}_2 \dfrac{(-i)^2}{2} \langle0|[[\hat{\tilde{\zeta}}^{(0)}_k(\tilde{\eta})\hat{\tilde{\zeta}}^{(0)}_{k'}(\tilde{\eta}),H_{\tilde{\zeta}} (\tilde{\eta}_1)],H_{\tilde{\zeta}} (\tilde{\eta}_2)]|0\rangle\nonumber\\
		=&~ {\cal P}_k^{\tilde{\zeta}(0)} - \int_{\tilde{\eta}_{min,1}}^{\tilde{\eta}}d\tilde{\eta}_1\int_{\tilde{\eta}_{min,2}}^{\tilde{\eta}_1} d\tilde{\eta}_2 \langle0|[[\hat{\tilde{\zeta}}^{(0)}_k(\tilde{\eta})\hat{\tilde{\zeta}}^{(0)}_{k'}(\tilde{\eta}),H_{\tilde{\zeta}}(\tilde{\eta}_1)],H_{\tilde{\zeta}}(\tilde{\eta}_2)]|0\rangle,
	\end{align}
	 where $\hat{\tilde{\zeta}}_k^{(0)}(\tilde{\eta})$ is defined as
		\begin{equation}
		\hat{\tilde{\zeta}}_k^{(0)}(\tilde{\eta}) = \tilde\zeta_k^{(0)}(\tilde{\eta})a({\bf k})+\tilde\zeta_k^{(0)\ast }(\tilde{\eta}) a^{\dagger}(-{\bf k}).
	\end{equation} 
	Here, $\tilde\zeta_k^{(0)}(\tilde{\eta})$ is the standard BD vacuum state for the non-canonical scalar field, whose formula has been shown in Eq. \eqref{BD}. In addition, $a^{\dagger}(-{\bf k})$ and $a({\bf k})$ are the creation and annihilation operators, respectively, which satisfy the following commutation relations as $[a({\bf k}),a^{\dagger}(-{\bf k'})] =\delta^3 ({\bf k}+{\bf k'})$, $[a({\bf k}),a({\bf k})]=0$, and $[a^{\dagger}({\bf k}),a^{\dagger}({\bf k})]=0$.

	What we need to define for now is the interacting Hamiltonian $H_{\tilde{\zeta}}$ in the above equation in order to obtain the correction to the power spectrum arising from the non-minimal coupling term $-h^2(\phi)H^2/12$. To do this, we first define the corresponding tree-level interacting Lagrangian as follows \cite{Ohashi:2013mka,Ohashi:2013qba}
	\begin{align}\label{App1}
		\mathcal{L}_{int} &= - \dfrac{a^4}{12}\left(\langle h^2\rangle+\dfrac{\partial\langle h^2\rangle}{\partial \phi}\delta\phi\right)\left(\langle H_{\mu\nu\lambda}\rangle+\delta H_{\mu\nu\lambda}\right)\left(\langle H^{\mu\nu\lambda}\rangle+\delta H^{\mu\nu\lambda}\right)\nonumber\\
		& \simeq a^4 E_{yz}\left(2\delta E_{yz}\tilde{\zeta}-\delta_{xz}h_{xy}-\delta E_{yz}h_{zz}+\delta E_{xy}h_{xx}\right),
	\end{align}
	here we have only kept the second-order terms of perturbations.
	 Next, we apply the Fourier transform to $\delta E_{ij} (\bf{x}, \tilde{\eta})$ such as \cite{Ohashi:2013mka,Ohashi:2013qba}
	 	\begin{equation}\label{}
		\delta E_{ij}({\bf x},\tilde{\eta}) = \int \dfrac{d^3k}{(2\pi)^{3/2}} e^{i{\bf k}\cdot {\bf x}}\delta \mathcal{E}_{ij}({\bf k},\tilde{\eta}),
	\end{equation}
	As a result, we can obtain a solution in the super-Hubble regime $(|k\tilde{\eta}| \ll 1)$ as \cite{Ohashi:2013mka,Ohashi:2013qba}	
	\begin{equation}\label{Eij}
		\delta\mathcal{E}_{ij}({\bf k},\tilde{\eta})=\dfrac{3H^2}{\sqrt{k^3}}\left[b({\bf k})+b^{\dagger}(-{\bf k})\right]\epsilon_{ij}({\bf k}),
	\end{equation}
	where  $b^{\dagger}(-{\bf k})$ and $b({\bf k})$ are creation and annihilation  operators, respectively, which satisfy the following commutation relations as $[b({\bf k}),b^{\dagger}(-{\bf k'})] =\delta^3 ({\bf k}+{\bf k'})$, $[b({\bf k}),b({\bf k})]=0$, and $[b^{\dagger}({\bf k}),b^{\dagger}({\bf k})]=0$. In addition, $\epsilon_{ij}({\bf k}) \equiv i\epsilon_{ijl}k^l/(\sqrt{2}k)$ is the polarization tensor, whose explicit components can be found in Ref.  \cite{Ohashi:2013qba}. 
	Consequently, the interacting Hamiltonian $H_{\tilde{\zeta}}$ can be defined to be  \cite{Ohashi:2013mka,Ohashi:2013qba}
	\begin{align}\label{2.18}
		H_{\tilde{\zeta}} (\tilde{\eta}) = -\dfrac{2E_{yz}}{H^4\tilde{\eta}^4}\int d^3k\delta\mathcal{E}_{yz}({\bf k},\tilde{\eta})\hat{\tilde{\zeta}}^{(0)}(-{\bf k},\tilde{\eta}),
		\end{align}
according to the relation, $ H_{\tilde{\zeta}} =-\int d^3x \mathcal{L}_{ {\tilde \zeta}}$, where $\mathcal{L}_{ {\tilde \zeta}}$ is nothing but the first term in Eq. \eqref{App1}. Consequently, we are able to define 
	\begin{align}
		\quad\delta\langle0|\hat{\tilde{\zeta}}_{k,\tilde{\eta}}\hat{\tilde{\zeta}}_{k',\tilde{\eta}}|0\rangle=&~{\cal P}_k^{\tilde{\zeta}}(\tilde{\eta},k)-{\cal P}_k^{\tilde{\zeta}(0)}(\tilde{\eta},k)\nonumber\\
		&\simeq -\dfrac{1}{2} \int_{\tilde{\eta}_{min,1}}^{\tilde{\eta}}d\tilde{\eta}_1\int_{\tilde{\eta}_{min,2}}^{\tilde{\eta}_1} d\tilde{\eta}_2 \langle0|[[\hat{\tilde{\zeta}}^{(0)}_k(\tilde{\eta})\hat{\tilde{\zeta}}^{(0)}_{k'}(\tilde{\eta}),H_{\tilde{\zeta}} (\tilde{\eta}_1)],H_{\tilde{\zeta}}(\tilde{\eta}_2)]|0\rangle\nonumber\\
		=&-\dfrac{1}{2}\int_{\tilde{\eta}_{min,1}}^{\tilde{\eta}}d\tilde{\eta}_1\int_{\tilde{\eta}_{min,2}}^{\tilde{\eta}_1}d\tilde{\eta}_2 \left(\dfrac{4E^2_{yz}}{H^8}\right)\int d^3k_1\int d^3 k_2 \nonumber\\
		& \times \dfrac{1}{\tilde{\eta}_1^4\tilde{\eta}_2^4} \langle0|[\hat{\tilde{\zeta}}_{k,\tilde{\eta}}^{(0)}\hat{\tilde{\zeta}}_{k',\tilde{\eta}}^{(0)}\hat{\tilde{\zeta}}_{k_1,\tilde{\eta}_1}^{(0)}-\hat{\tilde{\zeta}}^{(0)}_{k_1,\tilde{\eta}_1}\hat{\tilde{\zeta}}^{(0)}_{k,\tilde{\eta}}\hat{\tilde{\zeta}}^{(0)}_{k',\tilde{\eta}},\hat{\tilde{\zeta}}^{(0)}_{k_2,\tilde{\eta}_2}]\delta\mathcal{E}_{yz}(\tilde{\eta}_1,{\bf k}_1)\delta\mathcal{E}_{yz}(\tilde{\eta}_2,{\bf k}_2)|0\rangle\nonumber\\	
		=& -\dfrac{2E_{yz}^2}{H^8} \int_{\tilde{\eta}_{min,1}}^{\tilde{\eta}}\int_{\tilde{\eta}_{min,2}}^{\tilde{\eta}_1}d\tilde{\eta}_2 \int d^3 k_1 \int d^3 k_2 \nonumber\\
		& \times \dfrac{1}{\tilde{\eta}_1^4\tilde{\eta}_2^4} \langle0|\left[\hat{\tilde{\zeta}}^{(0)}_{k,\tilde{\eta}}[\hat{\tilde{\zeta}}^{(0)}_{k',\tilde{\eta}},\hat{\tilde{\zeta}}^{(0)}_{k_1,\tilde{\eta}_1}]+[\hat{\tilde{\zeta}}_{k,\tilde{\eta}}^{(0)},\hat{\tilde{\zeta}}_{k_1,\tilde{\eta}_1}^{(0)}]\hat{\tilde{\zeta}}^{(0)}_{k',\tilde{\eta}},\hat{\tilde{\zeta}}^{(0)}_{k_2,\tilde{\eta}_2}\right]\delta\mathcal{E}_{yz}(\tilde{\eta}_1,{\bf k}_1)\delta\mathcal{E}_{yz}(\tilde{\eta}_2,{\bf k}_2)|0\rangle.
	\end{align}
	Now, using the commutation relation,
	\begin{equation}\label{}
		\left[\hat{\tilde{\zeta}}^{(0)}(\tilde{\eta},{\bf k}),\hat{\tilde{\zeta}}^{(0)}(\tilde{\eta}',{\bf k}')\right] \simeq -\dfrac{i H^2 c_s^2}{6 \epsilon M_p^2}(\tilde{\eta}^3-\tilde{\eta}'^3)\delta^3({\bf k}+{\bf k}'),
	\end{equation}
	 we obtain
	\begin{align}
		&\delta\langle0|\hat{\tilde{\zeta}}_{k,\tilde{\eta}}\hat{\tilde{\zeta}}_{k',\tilde{\eta}}|0\rangle\nonumber\\
		& \simeq \dfrac{2i E_{yz}^2}{H^8} \dfrac{H^2c_s^2}{6\epsilon M_p^2} \int_{\tilde{\eta}_{min,1}}^{\tilde{\eta}}d\tilde{\eta}_1\int_{\tilde{\eta}_{min,2}}^{\tilde{\eta}_1}d\tilde{\eta}_2 \int d^3k_2 \dfrac{1}{\tilde{\eta}_1^4\tilde{\eta}_2^4}\langle0|[2(\tilde{\eta}^3-\tilde{\eta}_1^3)\hat{\tilde{\zeta}}^{(0)}_{k',\tilde{\eta}},\hat{\tilde{\zeta}}^{(0)}_{k_2,\tilde{\eta}_2}]\delta\mathcal{E}_{yz}(\tilde{\eta}_1,{\bf k})\delta\mathcal{E}_{yz}(\tilde{\eta}_2,{\bf k}_2)|0\rangle\nonumber\\
		&=\dfrac{4 E^2_{yz}}{H^8}\left(\dfrac{H^2c_s^2}{6\epsilon M_p^2}\right)^2\int_{\tilde{\eta}_{min,1}}^{\tilde{\eta}}d\tilde{\eta}_1\int_{\tilde{\eta}_{min,2}}^{\tilde{\eta}_1}d\tilde{\eta}_2\dfrac{1}{\tilde{\eta}_1^4\tilde{\eta}_2^4}(\tilde{\eta}^3-\tilde{\eta}_1^3)(\tilde{\eta}^3-\tilde{\eta}_2^2)\langle0|\delta\mathcal{E}_{yz}(\tilde{\eta}_1,{\bf k})\delta\mathcal{E}_{yz}(\tilde{\eta}_2,{\bf k}')|0\rangle\nonumber\\
		&=\dfrac{E_{yz}^2c_s^4}{9 H^4 \epsilon^2M_p^4}\int_{\tilde{\eta}_{min,1}}^{\tilde{\eta}}d\tilde{\eta}_1\dfrac{\tilde{\eta}^3-\tilde{\eta}_1^3}{\tilde{\eta}_1^4}\int_{\tilde{\eta}_{min,2}}^{\tilde{\eta}_1}d\tilde{\eta}_2\dfrac{\tilde{\eta}^3-\tilde{\eta}_2^3}{\tilde{\eta}_2^4}\langle0|\delta\mathcal{E}_{yz}(\tilde{\eta}_1,{\bf k})\delta\mathcal{E}_{yz}(\tilde{\eta}_2,{\bf k}')|0\rangle,
	\end{align}
	which will be identical to that defined in the case of canonical scalar field \cite{Ohashi:2013mka,Ohashi:2013qba} if we take limit $c_s \to 1$.
	
Thanks to the expression \eqref{Eij}, which can be written explicitly as \cite{Ohashi:2013qba}
	\begin{align}
		\delta\mathcal{E}_{yz}(\tilde{\eta}_1,{\bf k}) &= \dfrac{3H^2}{\sqrt{k^3}}\left[b({\bf k})+b^{\dagger}(-{\bf k})\right]\dfrac{i}{\sqrt{2}}\cos\theta,
	\end{align}
	 we are able to define the following formula,
	\begin{align}
		\langle0|\delta\mathcal{E}_{yz}(\tilde{\eta}_1,{\bf k})\delta\mathcal{E}_{yz}(\tilde{\eta}_2,{\bf k}')|0\rangle &= \langle0|-\dfrac{9H^4}{2\sqrt{k^3k'^3}}\left[b({\bf k})+b^{\dagger}(-{\bf k})\right]\left[b({\bf k}')+b^{\dagger}(-{\bf k}')\right](-\cos^2\theta)|0\rangle\nonumber\\
		&= \dfrac{9H^4}{2k^3}\delta^3({\bf k}+{\bf k}')\cos^2\theta.
	\end{align}
	And thus we have
	\begin{align}
		\delta\langle0|\hat{\tilde{\zeta}}_{k,\tilde{\eta}}\hat{\tilde{\zeta}}_{k',\tilde{\eta}}|0\rangle = \dfrac{E_{yz}^2c_s^4}{2k^3 \epsilon^2M_p^4}\delta^3({\bf k}+{\bf k}')\cos^2\theta\int_{-1/(c_s k)}^{\tilde{\eta}}d\tilde{\eta}_1\dfrac{\tilde{\eta}^3-\tilde{\eta}_1^3}{\tilde{\eta}_1^4}\int_{-1/(c_s k')}^{\tilde{\eta}_1}d\tilde{\eta}_2\dfrac{\tilde{\eta}^3-\tilde{\eta}_2^3}{\tilde{\eta}_2^4}.
	\end{align}
	The integral in the above equation can be calculated as follows \cite{Do:2020ler,Ohashi:2013mka}
	\begin{align}
		\int_{-1/(c_s k)}^{\tilde{\eta}}d\tilde{\eta}_1\dfrac{\tilde{\eta}^3-\tilde{\eta}_1^3}{\tilde{\eta}_1^4}\int_{-1/(c_s k')}^{\tilde{\eta}_1}d\tilde{\eta}_2\dfrac{\tilde{\eta}^3-\tilde{\eta}_2^3}{\tilde{\eta}_2^4} \simeq \dfrac{1}{2} N_{c_s k}^2,
	\end{align}
	here $N_{c_s k }\equiv \ln |\tilde{\eta} c_s k|$ is the e-fold number and the approximation $|\tilde{\eta} c_s k|\ll 1$ has been used. 
	Thus, we have
	\begin{equation}\label{}
		\delta\langle0|\hat{\tilde{\zeta}}_{k,\tilde{\eta}}\hat{\tilde{\zeta}}_{k',\tilde{\eta}}|0\rangle =  \dfrac{E_{yz}^2c_s^4 N_{c_s k}^2\cos^2\theta}{4 \pi^2 \epsilon^2 M_p^4 }\dfrac{2\pi^2}{k^3}\delta^3({\bf k}+{\bf k}'),
	\end{equation} 
	which implies that
	\begin{align}
		{\cal P}^{\tilde{\zeta}}_{k,\text{ani}} ({{\bf k}})  \equiv{\cal P}_k^{\tilde{\zeta}(0)} +\delta{\cal P}_k^{\tilde{\zeta}} =   {\cal P}_k^{\tilde{\zeta}(0)} \left(1+\dfrac{2c_s^5 E_{yz}^2 N_{c_sk}^2}{H^2 M_p^2\epsilon}\cos^2\theta\right).
	\end{align}

\end{document}